\documentclass[11pt]{article}

\usepackage[final]{acl}
\usepackage{amsmath}
\usepackage{times}
\usepackage{latexsym}
\usepackage{CJKutf8}

\usepackage[T1]{fontenc}

\usepackage[utf8]{inputenc}

\usepackage{microtype}
\usepackage{siunitx}

\usepackage{inconsolata}

\usepackage{graphicx}
\usepackage{booktabs}
\usepackage{float}

\title{Evaluating Reliability Asymmetries in Chinese Factual Search and AI Answers}

\author{
  Geng Liu\textsuperscript{1} \quad
  Li Feng\textsuperscript{2} \quad
  Mengxiao Zhu\textsuperscript{2} \quad
  Francesco Pierri\textsuperscript{1}\footnotemark[1]\thanks{Corresponding author. Email: \texttt{francesco.pierri@polimi.it}} \\
  \textsuperscript{1}Department of Electronics, Information and Bioengineering, Politecnico di Milano, Milan, Italy \\
  \textsuperscript{2}University of Science and Technology of China, Hefei, China \\
  \texttt{\{geng.liu,francesco.pierri\}@polimi.it} \\
  \texttt{fengli@mail.ustc.edu.cn, mxzhu@ustc.edu.cn}
}

\begin{document}
\maketitle

\begin{abstract}
Search engines and AI-powered systems increasingly mediate access to factual information, yet their reliability remains difficult to evaluate in realistic information-seeking settings. 
We study this problem in the Chinese web ecosystem by constructing a query-based fact-checking dataset from real Chinese search logs and comparing nine systems across traditional search engines, standalone large language models, and search-integrated AI Overviews. 
Focusing on factual Chinese-language factual Yes/No questions, we evaluate whether systems provide correct, incorrect, or uncertain decisions against evidence-derived ground truth. 
We find that systems are similarly accurate when they provide definitive answers, but differ sharply in how often they do so. 
Conditional accuracy ranges from 73.2\% to 78.9\%, yet search engines answer definitively on over 83\% of queries, while Qwen-Max does so on fewer than half. 
We also find a consistent polarity gap: all systems perform better on yes-labeled queries than on no-labeled queries.
We also use Baidu Index data to identify Chinese provinces with higher health-related search attention, which may indicate greater potential exposure to misinformation. Overall, our results show that reliability depends not only on whether systems are correct when they answer, but also on how often they answer, how they handle negative claims, and where information demand may increase exposure risks.
\end{abstract}

\section{Introduction}

Online search engines have long shaped how users access factual information across domains such as health, education, and public affairs~\citep{bachl2024search,ye2020investigating}.
These systems are now increasingly complemented by large language models (LLMs) and AI-generated search summaries that provide direct natural-language answers rather than only ranked links~\citep{yang2025search+,liu-etal-2023-evaluating,10.1145/3715275.3732089}.
This shift matters because users who receive authoritative-sounding answers may be exposed to inaccurate, unsupported, or overly uncertain information~\citep{liu-etal-2023-evaluating,10.1145/3715275.3732089}.
A substantial body of work has addressed automated fact-checking, typically combining claim identification, evidence retrieval, and veracity assessment~\citep{nakov2021automated,guo-etal-2022-survey,li-etal-2024-self}.
Although recent efforts have introduced Chinese resources for evidence-based fact-checking and passage ranking~\citep{hu-etal-2022-chef,lin2024cfever,xie2023t2ranking}, these resources are not designed to evaluate the factual reliability of user-facing information-access systems.
Unified evaluations on real Chinese queries remain limited, especially those comparing traditional search engines, LLMs, and search-integrated AI Overviews under the same framework~\citep{thorne-etal-2018-fever,jin-etal-2019-pubmedqa,liu-etal-2023-evaluating,10.1145/3715275.3732089,fernandez2025evaluating}.
Because these systems increasingly mediate factual information seeking, evaluating their outputs on real user queries offers a way to approximate the reliability of information that Chinese users may encounter in everyday search and AI-assisted information access.

In this work, we evaluate factual reliability in Chinese information-access systems using real search-log queries and evidence-derived ground truth.
We focus on factual Yes/No questions because they allow system outputs to be mapped into \textit{Yes}, \textit{No}, or \textit{Uncertain} decisions against verifiable evidence.
Specifically, we ask:
\begin{itemize}
    \item[\textbf{RQ}] How reliable are Chinese information-access systems on real factual user queries?
\end{itemize}
To answer this question, we construct a query-based fact-checking dataset from real Chinese search logs and evaluate nine systems spanning three paradigms: traditional search engines, LLMs, and search-integrated AI Overviews. As a secondary contextual analysis, we use population-normalized Baidu Index to map geographic variation in health-related search attention across Chinese provinces.

Our contributions are as follows:
\begin{itemize}
    \item We introduce an evidence-derived benchmark of Chinese factual Yes/No queries from real search logs, enabling unified evaluation of heterogeneous user-facing information-access systems.
    \item We compare nine systems across search engines, LLMs, and AI Overviews, showing that reliability differences are driven not only by conditional accuracy but also by large variation in decision rate.
    \item We identify a consistent polarity asymmetry across paradigms: systems are substantially less reliable on no-labeled queries than on yes-labeled queries, suggesting that aggregate accuracy can obscure systematic weaknesses in how systems handle queries whose evidence-derived answer is \textit{No}.
\end{itemize}

\begin{figure*}
    \centering
    \includegraphics[width=\textwidth]{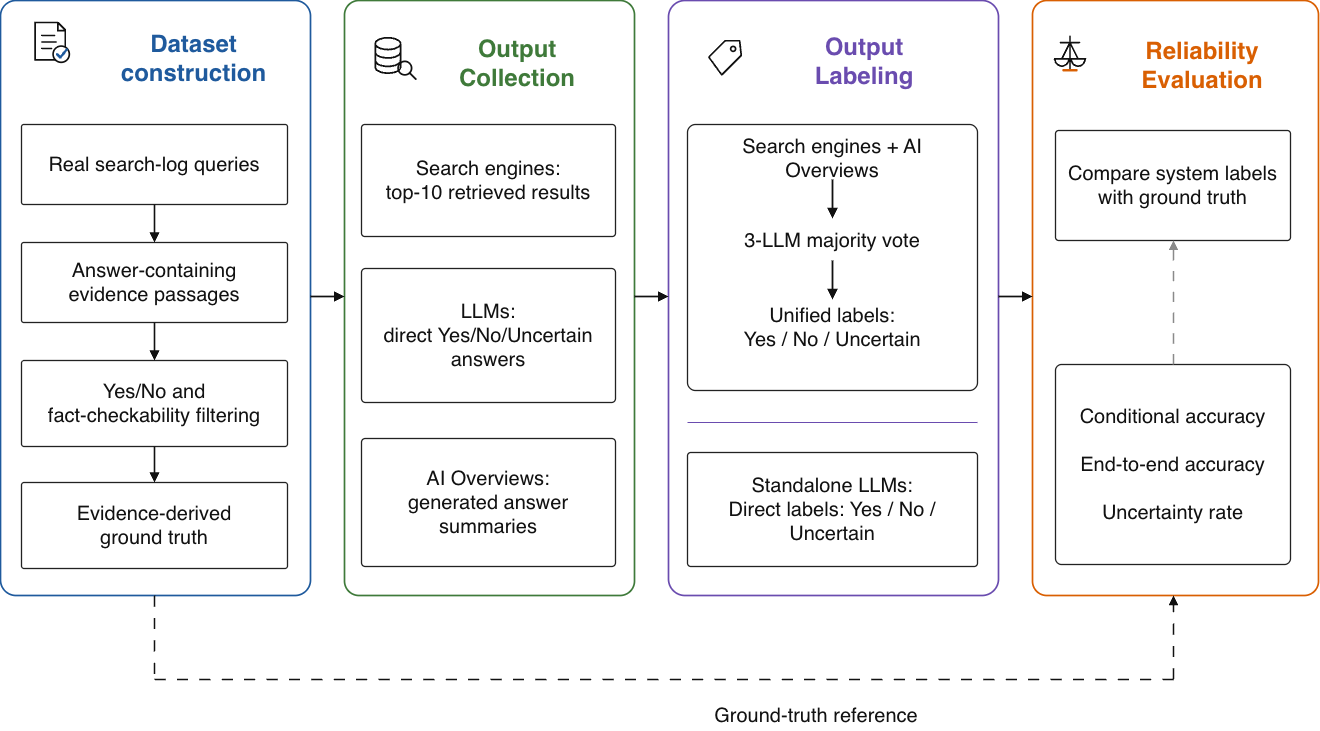}
    \caption{Workflow for our research pipeline.}
    \label{fig:pipeline}
\end{figure*}

\section{Related Work}

\subsection{Automated Fact-Checking}

Automated fact-checking is traditionally framed as a pipeline of claim detection, evidence retrieval, and veracity prediction. This paradigm was consolidated with datasets such as FEVER~\cite{thorne-etal-2018-fever}, and later extended with improved retrieval and entailment models~\cite{zhou-etal-2019-gear}. However, these systems rely on curated English corpora and controlled evidence sources, limiting their applicability to real-world user queries.

Recent work has explored using LLMs for fact-checking, either as veracity classifiers or as automatic annotators. While LLMs can achieve moderate agreement with human judgments, they often hallucinate~\cite{10.1145/3571730} and behave inconsistently, raising concerns about their reliability in realistic settings~\cite{nakov2021automated,lin-etal-2022-truthfulqa}. Building on these concerns, DeVerna et al.\ show that even when LLMs are equipped with explicit reasoning steps and web-search evidence, they still struggle to reliably verify political claims without curated context~\cite{deverna2025large}. Nelson et al.\ similarly report that traditional search engines outperform AI chatbots on factual reliability benchmarks~\cite{nelson2025reliability}. 
In the health domain, Fernandez et al.\ compare search engines and LLMs for answering health questions, finding that traditional search engines remain competitive despite the emergence of AI systems~\cite{fernandez2025evaluating}. 

\subsection{Search Queries and Public Information Demand}

Prior work has examined how search queries capture users' information needs and has introduced a range of topical taxonomies. Foundational analyses of Google search logs identify broad query categories such as health, entertainment, commerce, and navigation~\cite{chuang2003enriching}. More recent studies extend this line of work by analyzing domain-specific behaviors and the reliability of health-related search interactions~\cite{bach2020studying,rohatgi2021were,sun2024trusting,hashavit2024impact,bachl2024search}. In the Chinese context, related research proposes comparable taxonomies: Ye et al.\ examine Sogou search behavior during the COVID-19 pandemic~\cite{ye2020investigating}, and the T2Ranking dataset provides structured domain labels across education, healthcare, IT, and finance~\cite{xie2023t2ranking}.

Search logs and search-volume indices have also been used to study public attention and information demand. In China, Baidu Index provides regional and temporal signals of aggregate search interest, and prior studies have used it to track public responses during health crises~\cite{gong2020online,wang2022geographic}, environmental and societal risks, and mental health and well-being~\cite{tan2022using}. These signals should be interpreted as proxies for information-seeking behavior rather than direct measures of individual demand, exposure, or offline outcomes.

\subsection{Positioning of Our Work}
Our work connects automated fact-checking with online searches by evaluating Chinese language web systems on factual Yes/No queries derived from real search logs. 
Rather than measuring fact-checking performance on curated claims alone, we approximate the reliability of factual information that users may encounter when posing real queries to search engines, LLMs, and search-integrated AI Overviews. 
We treat these paradigms as different modalities through which users might search for information  rather than fully independent technical architectures: search engines expose retrieved evidence directly, LLMs answer from parametric knowledge, and AI Overviews combine search integration with generated summaries through opaque provider-specific pipelines~\cite{fan2024survey}.
We compare these paradigms under a unified framework, measuring both accuracy and decision rate.
We also use population-normalized Baidu Index as a secondary contextual signal to identify provinces with higher population-normalized health-related search attention, without estimating province-specific accuracy or user-level exposure.

\section{Data and Methods}

\subsection{Dataset Construction}
\label{sec:dataset-construction}

To evaluate factual correctness across information-access systems on real-world Chinese queries, we construct a query-based fact-checking dataset from real user search logs, as illustrated in Figure~\ref{fig:pipeline}.

\textbf{Raw data.} We build on T2Ranking~\cite{xie2023t2ranking}, which contains Chinese search queries and human-annotated passages. The original benchmark includes \num{307706} queries with four-level passage relevance labels: Level 0 indicates a complete mismatch between the query and passage, Level 1 indicates topical relevance without satisfying the information need, Level 2 indicates partial satisfaction of the information need, and Level 3 indicates that the passage satisfies the information need and contains the answer. We focus on the \num{135860} queries that have Level 3 passages, as they provide the most suitable evidence context for deriving query-level Yes/No labels.

\textbf{Query filtering.} We apply a two-step procedure. First, we retain only queries that can be answered with a binary \textit{Yes}/\textit{No} response, yielding \num{22052} queries. Second, among these, we retain only fact-checkable queries whose answers can be verified from evidence, excluding subjective or opinion-based cases. 
For each retained query, we provide the query and its associated evidence passage to an LLM annotator, which assigns one of three answer labels: \textit{Yes}, \textit{No}, or \textit{Uncertain}. The resulting label, rather than the original T2Ranking relevance label, serves as the evidence-derived ground-truth answer for the query.  A query-passage pair is labeled \textit{Uncertain} when the passage does not provide sufficient or relevant evidence to support either answer. Queries whose evidence passages are labeled \textit{Uncertain} are excluded from the final evaluation set. After filtering and evidence labeling, the final benchmark contains \num{12165} factual Yes/No queries, including \num{8385} labeled as \textit{Yes} and \num{3780} labeled as \textit{No}. 

The whole procedure is assisted by DeepSeek-R1 via the DeepSeek API,\footnote{\url{https://api-docs.deepseek.com/}.} using the prompt templates reported in Appendix~\ref{appendix:yes_no_question}, Appendix~\ref{appendix:fact_checking} and Appendix~\ref{appendix:answer_labeling_prompt}. Details on a manual evaluation of this procedure are provided in a later paragraph.

\textbf{Topic definition.} To characterize the topical distribution of fact-checking queries, we adopt the T2Ranking taxonomy and refine it by splitting ``General'' into ``Society'' and ``Finance'', renaming ``E-commerce'' to ``Shopping'', and adding ``Adult''. The resulting taxonomy contains ten top-level domains; each query is then classified into one of these topics by DeepSeek-R1. Prompt details are reported in Appendix~\ref{appendix:topic_recognition}. Figure~\ref{fig:dataset-topic-distribution} shows the distribution of topics after filtering and annotation. The dataset is dominated by Health queries, followed by Technology and Finance, while several smaller categories contain substantially fewer instances. 

\begin{figure}[t]
    \centering
    \includegraphics[width=1\linewidth]{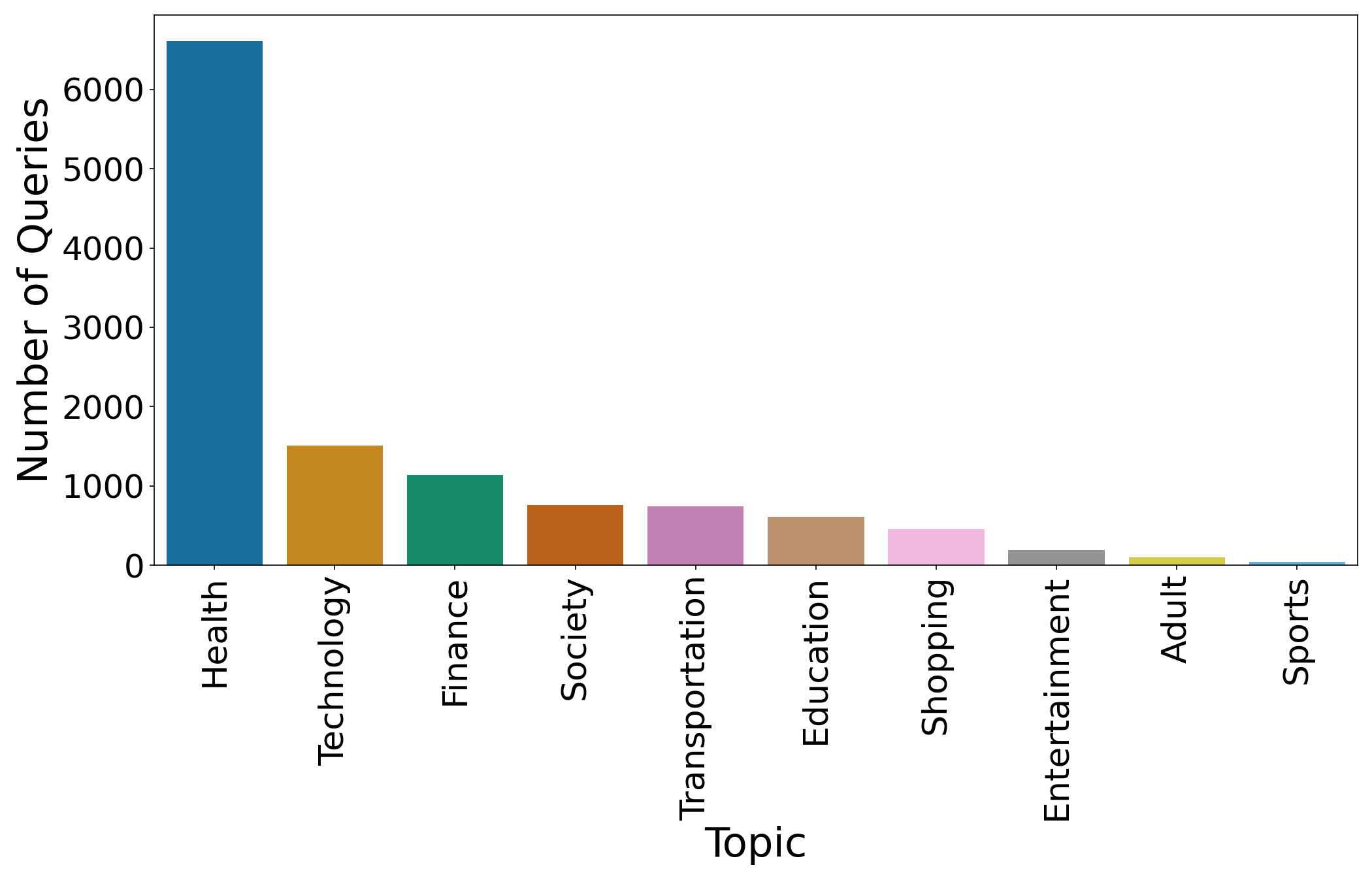}
    \caption{Topic distribution of the fact-checking query dataset after filtering and annotation.}
    \label{fig:dataset-topic-distribution}
\end{figure}

\textbf{Manual validation of the LLM-based preprocessing.} Since our analysis relies on an LLM-assisted pipeline to construct the benchmark, we manually validate random samples from each preprocessing stage before evaluating systems' fact-checking accuracy. Three annotators independently review 100 instances for Yes/No identification, fact-checkability identification, passage answer labeling and topic annotation. Annotators reach a non-tied majority decision for 100.0\% of Yes/No identification cases, 100.0\% of fact-checkability cases, and 97.0\% of passage answer-labeling cases. For topic labeling, the LLM-assigned topic matches the human reference in 95.0\% of sampled cases, and three-annotator agreement is high (Fleiss' $\kappa=0.919$). These validation results suggest that the benchmark construction procedure provides a sufficiently stable empirical basis for the system-level comparisons below.

\subsection{Information Systems and Evaluation Pipeline}

We evaluate nine systems spanning three paradigms: search engines (Baidu, Bing China, Sogou), LLMs (Qwen-Max, GLM-4-Plus, DeepSeek-R1), and search-integrated AI Overviews (Baidu, Bing, Sogou). For search engines, which return ranked result lists, we evaluate under two simulated user behaviors (lazy and diligent); for LLMs and AI Overviews, which produce single outputs, we evaluate directly. We use these categories to distinguish user-facing modes of information access rather than mutually exclusive technical architectures. Search engines expose ranked web evidence, LLMs return direct answers without external retrieval, and AI Overviews provide generated summaries through opaque provider-specific pipelines. We therefore evaluate all systems as black-box interfaces and compare user-facing decisions under a common label space.

All systems are tested on the same Chinese-language Yes/No factual-query set from Section~\ref{sec:dataset-construction}. For search engines and AI Overviews, we query web interfaces in real time and collect Chinese-language outputs between January and March 2026. For search engines, we retrieve the top 10 results per query via Firecrawl\footnote{\url{https://www.firecrawl.dev/}}; for AI Overviews, we capture the generated summary text and associated citations. For LLMs, we use zero-shot querying without external retrieval via official APIs\footnote{\url{https://www.alibabacloud.com/help/en/model-studio/qwen-api-via-dashscope}, \url{https://docs.bigmodel.cn/cn/api/introduction}, \url{https://api-docs.deepseek.com/}.} with temperature set to 0.01. All LLMs were evaluated using the default inference configuration provided by their official APIs at the time of data collection. The corresponding prompt templates are reported in Appendix~\ref{app:prompts-bilingual}.

\textbf{Output annotation.} The systems we evaluate do not return the same type of output: search engines return ranked web results, AI Overviews return generated summaries, and LLMs return direct answers. We therefore annotate different units for each system type before comparing them with the evidence-derived ground truth. For search engines, the annotation unit is an individual retrieved result. For each query, we retrieve up to 10 results, extract the textual content of each result, and label each result separately as \textit{Yes}, \textit{No}, or \textit{Uncertain}. These result-level labels are later aggregated into query-level decisions by the lazy and diligent user models described below.

AI Overviews return a single generated summary for a query when such a module is available. We label each collected summary once using the same \textit{Yes}/\textit{No}/\textit{Uncertain} label space. If no AI Overview is returned, the case is recorded as missing in the coverage analysis and is not passed to the classifier ensemble. Relative to the \num{12165}-query benchmark, usable AI Overview outputs were available for \num{12125} Baidu queries, \num{12159} Bing queries, and \num{8454} Sogou queries, leaving \num{40}, \num{6}, and \num{3711} missing or unavailable cases, respectively. LLMs are evaluated differently: they are prompted directly to output \textit{Yes}, \textit{No}, or \textit{Uncertain}, and this native response is used as the system decision without additional classifier labeling.

For search-engine results and AI Overview summaries, labels are produced by an ensemble of three LLM classifiers: Qwen-Max, Ernie 4.5 turbo 32k, and DeepSeek-R1. Each classifier receives the query and the corresponding system output, predicts one label independently, and the final label is assigned by majority vote. If the three classifiers produce three different labels, we assign \textit{Uncertain}. This procedure follows recent work on LLM-assisted evaluation~\citep{zheng2023judging}; prompt templates are reported in Appendix~\ref{app:prompts-bilingual}. Because this step relies on automatic classifiers, we use independent classifier calls with majority voting and validate the assigned labels against human annotations on sampled query-system pairs.

\textbf{Validation of output annotation.} We validate this annotation layer against human annotations on \num{600} query-system pairs, covering \num{300} AI Overview outputs and \num{300} search-engine outputs. Three independent annotators label each sampled output, and disagreements are resolved into a final human-adjudicated reference label. For AI Overview outputs, annotator agreement is substantial (Fleiss' $\kappa=0.707$), and the LLM-majority label is closely aligned with the human reference (Cohen's $\kappa=0.749$; raw agreement = 0.860). For search-engine outputs, agreement is lower but remains informative, reflecting the greater ambiguity of judging heterogeneous open-web results (Fleiss' $\kappa=0.472$; Cohen's $\kappa=0.661$; raw agreement = 0.800). In particular, these results indicate that the annotation procedure is more reliable for concise AI-generated summaries than for ranked search results, while still providing a defensible measurement foundation for the comparative analyses reported below.

\subsection{User Models and Evaluation Metrics}

To evaluate search-engine performance from the user's perspective, we define two query-level decision models that simulate different information-seeking behaviors~\citep{bachl2024search}. The \textit{lazy} user examines ranked results sequentially and returns the first definitive \textit{Yes} or \textit{No} label encountered; if no such label is found, the query is classified as \textit{No decision}. The \textit{diligent} user collects the first three definitive \textit{Yes}/\textit{No} labels and applies majority voting, with queries receiving fewer than three definitive labels also marked as \textit{No decision}.

For AI Overviews and LLMs, the lazy/diligent distinction does not apply, since these systems return a single generated output rather than a ranked list. We therefore evaluate each output directly against the Yes/No ground truth, labeling responses that do not commit to a definitive answer as \textit{Uncertain}.

We report two complementary metrics over each system's usable-output set. Let $\mathcal{Q}$ denote the system-specific set of queries for which usable output was collected, and let $\mathcal{D} \subseteq \mathcal{Q}$ denote the subset receiving a definitive decision. We define:

\begin{align*}
\textit{Conditional Accuracy} &= \frac{|\{q \in \mathcal{D} : \text{correct}\}|}{|\mathcal{D}|} \\
\textit{End-to-end Accuracy} &= \frac{|\{q \in \mathcal{D} : \text{correct}\}|}{|\mathcal{Q}|} \\
\textit{Uncertainty Rate} &= \frac{|\mathcal{Q} \setminus \mathcal{D}|}{|\mathcal{Q}|}
\end{align*}

Conditional accuracy measures correctness among definitive answers, whereas end-to-end accuracy additionally penalizes uncertain or no-decision outputs within the usable-output set. For search engines, we report these metrics under both lazy and diligent user models. For AI Overviews and LLMs, we report the same metrics without user-model simulation. Missing or unavailable outputs are not included in $\mathcal{Q}$; the results should therefore be interpreted as available-output measurements rather than full benchmark coverage-adjusted accuracy.

\section{Results}

\subsection{Accuracy across systems}

\begin{figure}
    \centering
    \includegraphics[width=1\linewidth]{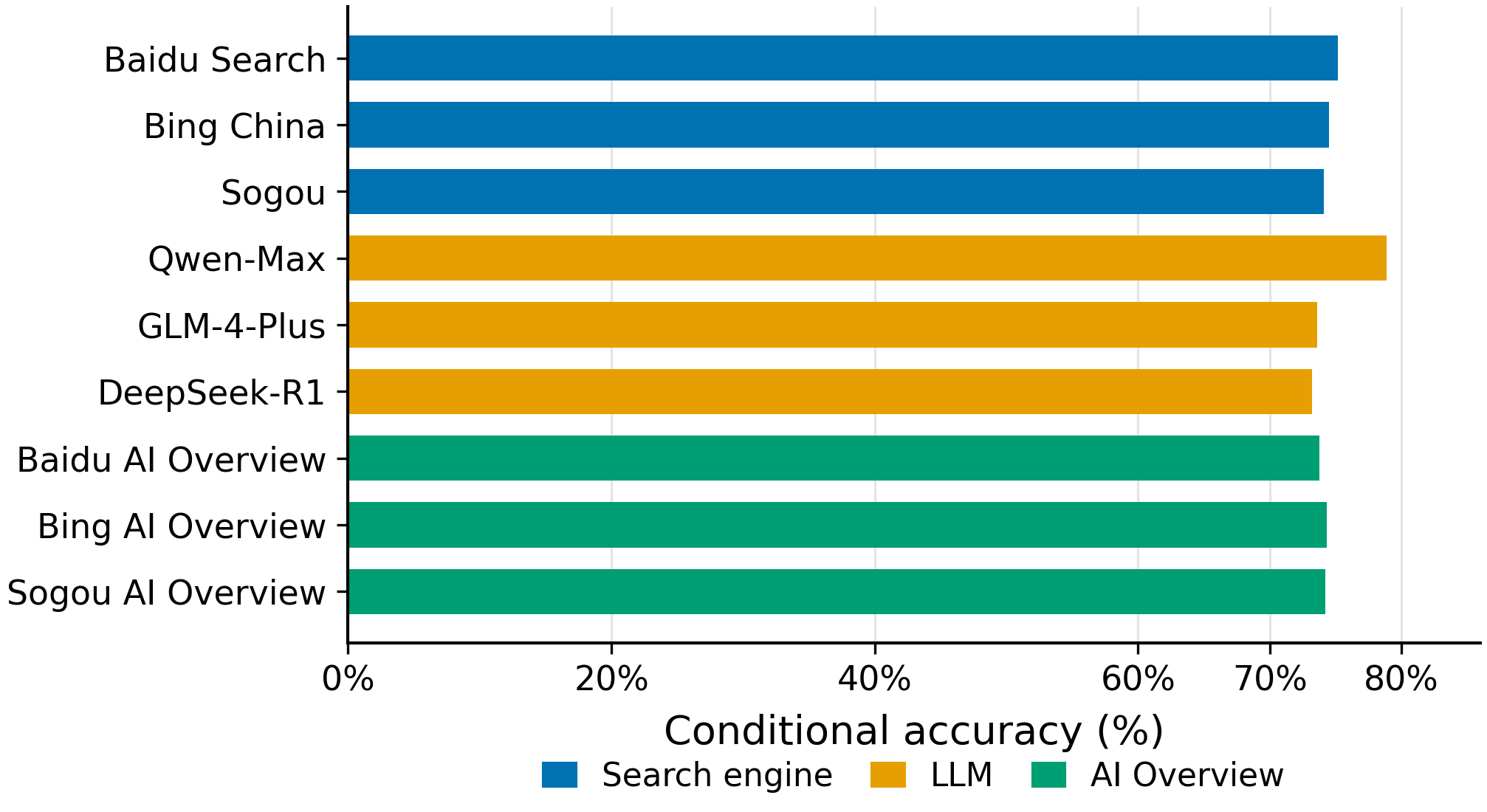}
    \caption{Conditional accuracy across search engines (lazy user), LLMs, and AI Overviews. Accuracy is computed on definitive outputs only, with uncertain or no-decision outputs excluded from the denominator. Systems are grouped by type.}
    \label{fig:accuracy-excluding-uncertain-systems}
\end{figure}

As shown in Figure~\ref{fig:accuracy-excluding-uncertain-systems}, conditional accuracy is broadly comparable across the nine systems, ranging from 73.2\% (DeepSeek-R1) to 78.9\% (Qwen-Max), with no system category clearly outperforming the others.

\begin{figure}[t]
    \centering
    \includegraphics[width=\linewidth]{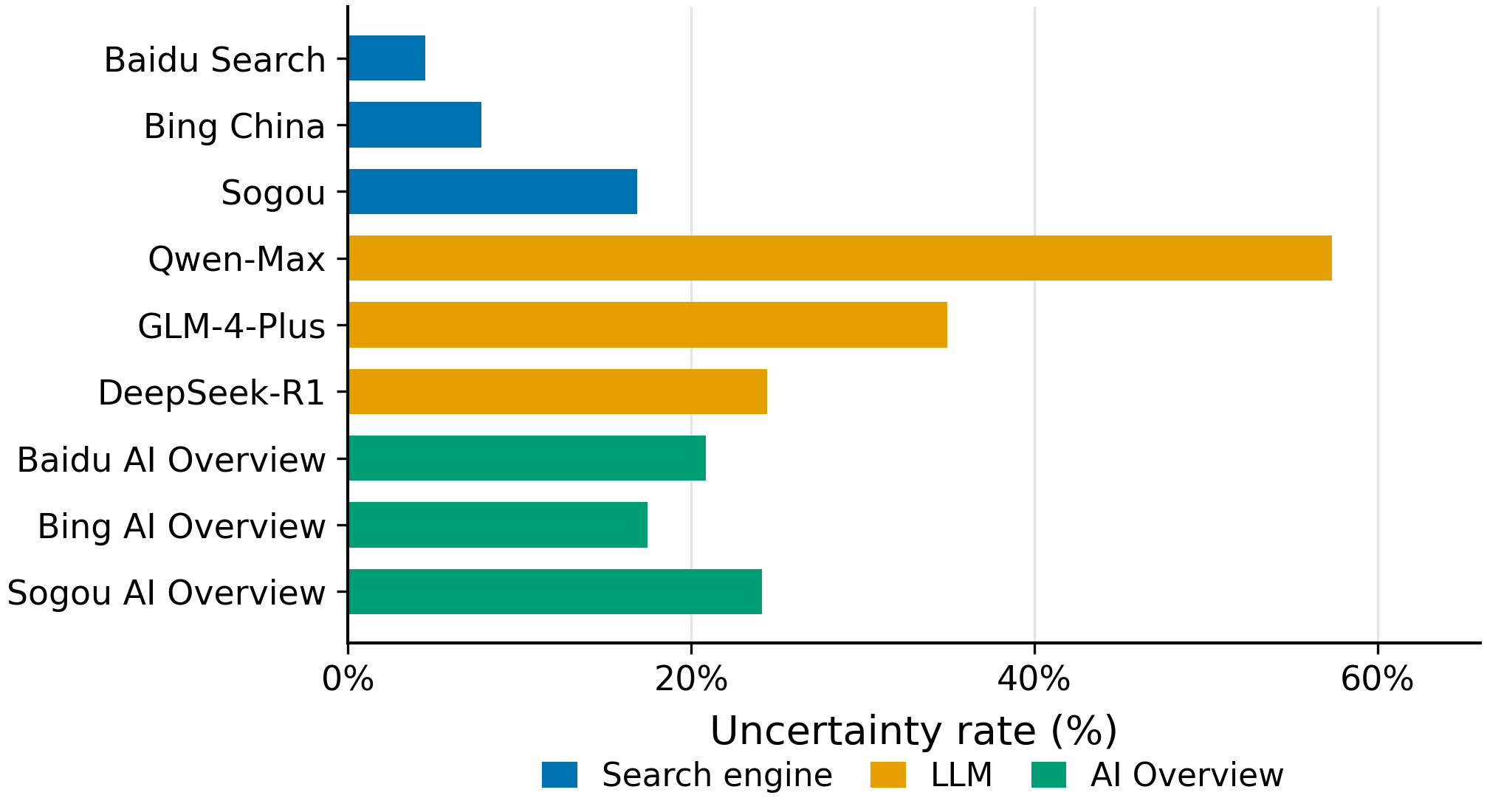}
    \caption{Uncertainty rates across search engines (lazy user), LLMs, and AI Overviews. The uncertainty rate is the proportion of \textit{No decision} or uncertain outcomes within each system's usable-output set. Systems are grouped by type.}
    \label{fig:uncertainty-rates-systems}
\end{figure}

However, systems differ substantially in how often they provide a definitive answer. As shown in Figure~\ref{fig:uncertainty-rates-systems}, search engines answer definitively in more than 83 out of every 100 queries, whereas Qwen-Max provides a definitive answer for fewer than half. AI Overviews fall between the two groups, producing uncertain outputs for roughly one in five usable outputs.

\begin{figure}[t]
    \centering
    \includegraphics[width=\linewidth]{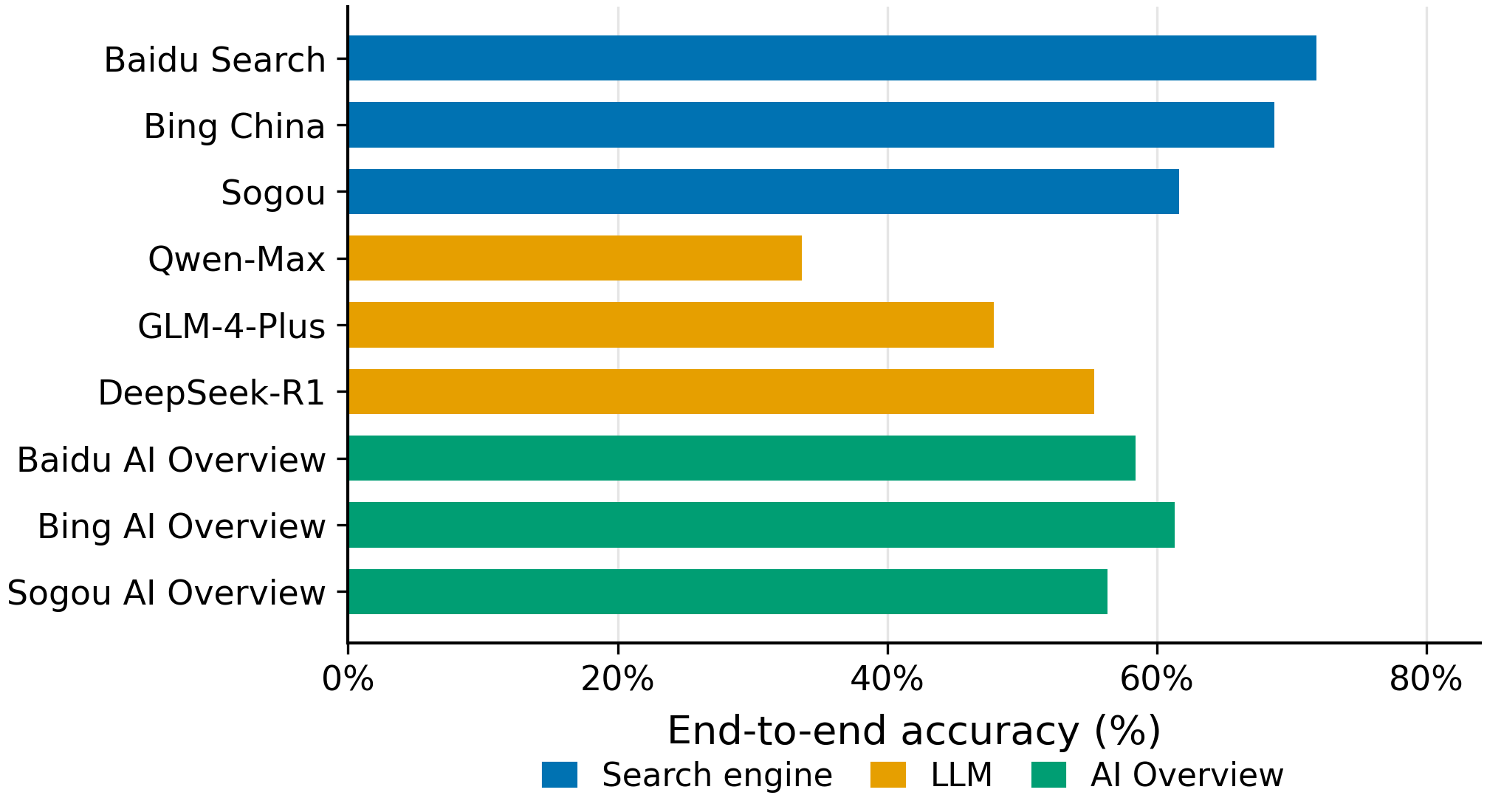}
    \caption{End-to-end accuracy across systems. Non-definitive outputs are counted as errors within each system's usable-output set. Search-engine results use the lazy user model.}
    \label{fig:end-to-end-accuracy-systems}
\end{figure}

As shown in Figure~\ref{fig:end-to-end-accuracy-systems}, system rankings change when uncertain or no-decision outputs are counted as errors. Search engines achieve end-to-end accuracy of 61.6\%--71.8\%, AI Overviews range from 56.3\% to 61.3\%, and LLMs range from 33.7\% to 55.3\%. Because coverage differs across web-facing systems, these estimates should be interpreted as available-output measurements rather than full benchmark coverage-adjusted accuracy. Qwen-Max illustrates this distinction: it has the highest conditional accuracy among definitive answers, but the lowest end-to-end accuracy because it provides a definitive answer for only 42.7\% of queries. These findings suggest that the main cross-system difference lies not in whether systems are correct once they answer, but in whether they answer at all. Appendix Table~\ref{tab:overall-performance} reports the corresponding numeric values.

\subsection{Accuracy of Search Engines}

\begin{figure}[t]
    \centering
    \includegraphics[width=1\linewidth]{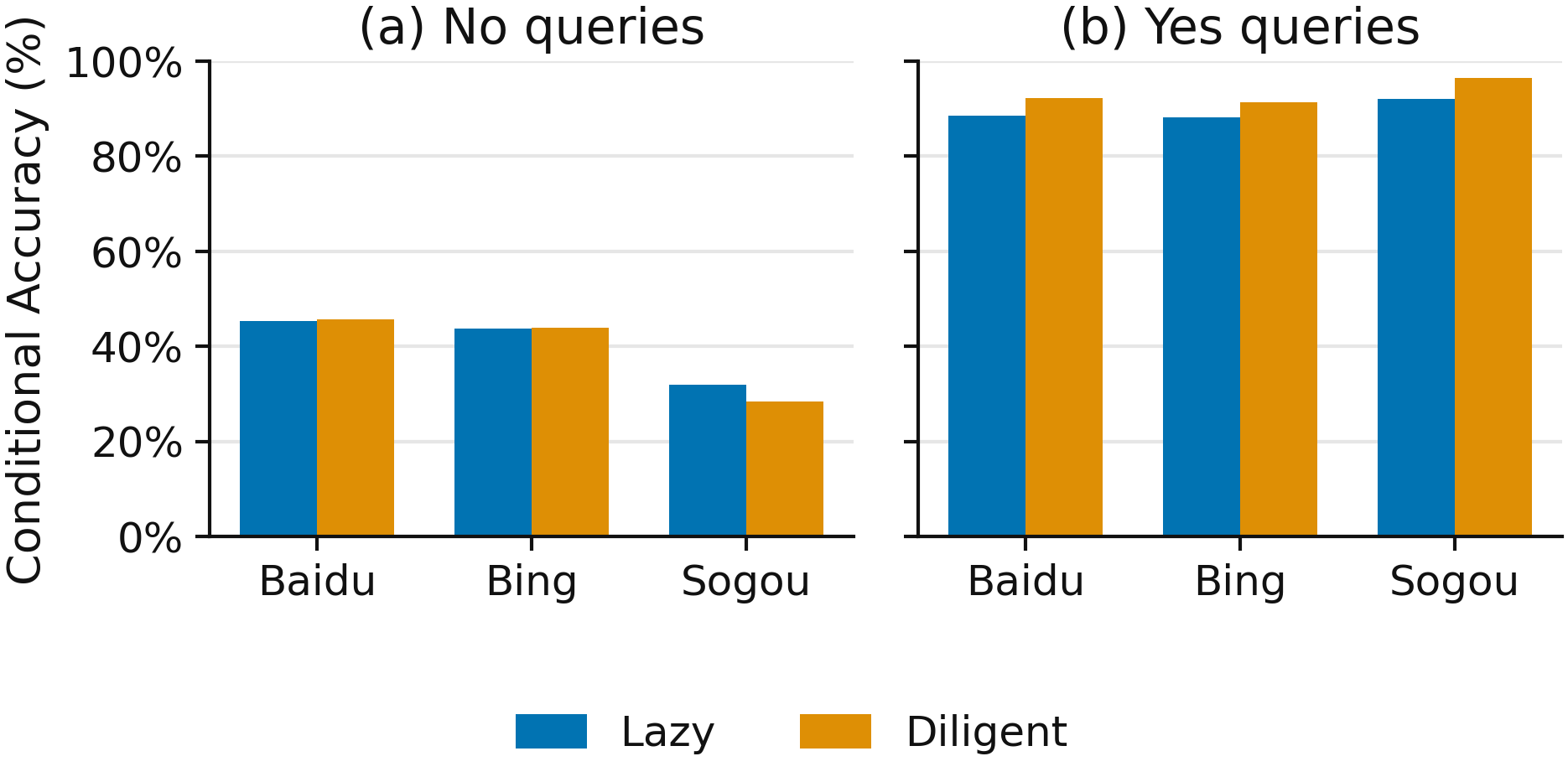}
    \caption{Search engine accuracy by answer polarity, excluding no-decision cases. Panels show (a) No and (b) Yes ground-truth queries; bar colors compare lazy and diligent user models.}
    \label{fig:se-yes-no-accuracy}
\end{figure}

As shown in Figure~\ref{fig:se-yes-no-accuracy}, we observe a persistent gap
between ``Yes'' and ``No'' queries across all three search engines. In the lazy
setting, accuracy on ``Yes'' queries exceeds 88\%, more than double the
32\%--45\% observed on ``No'' queries. The diligent setting raises accuracy on
``Yes'' queries further, above 91\%, but leaves accuracy on ``No'' queries
largely unchanged. This indicates a strong polarity asymmetry: search engines
are much more reliable when the correct answer is ``Yes'' than when the correct
answer is ``No'', and this gap persists even under the diligent user model.

\subsection{Accuracy of LLMs}

\begin{figure}[t]
    \centering
    \includegraphics[width=1\linewidth]{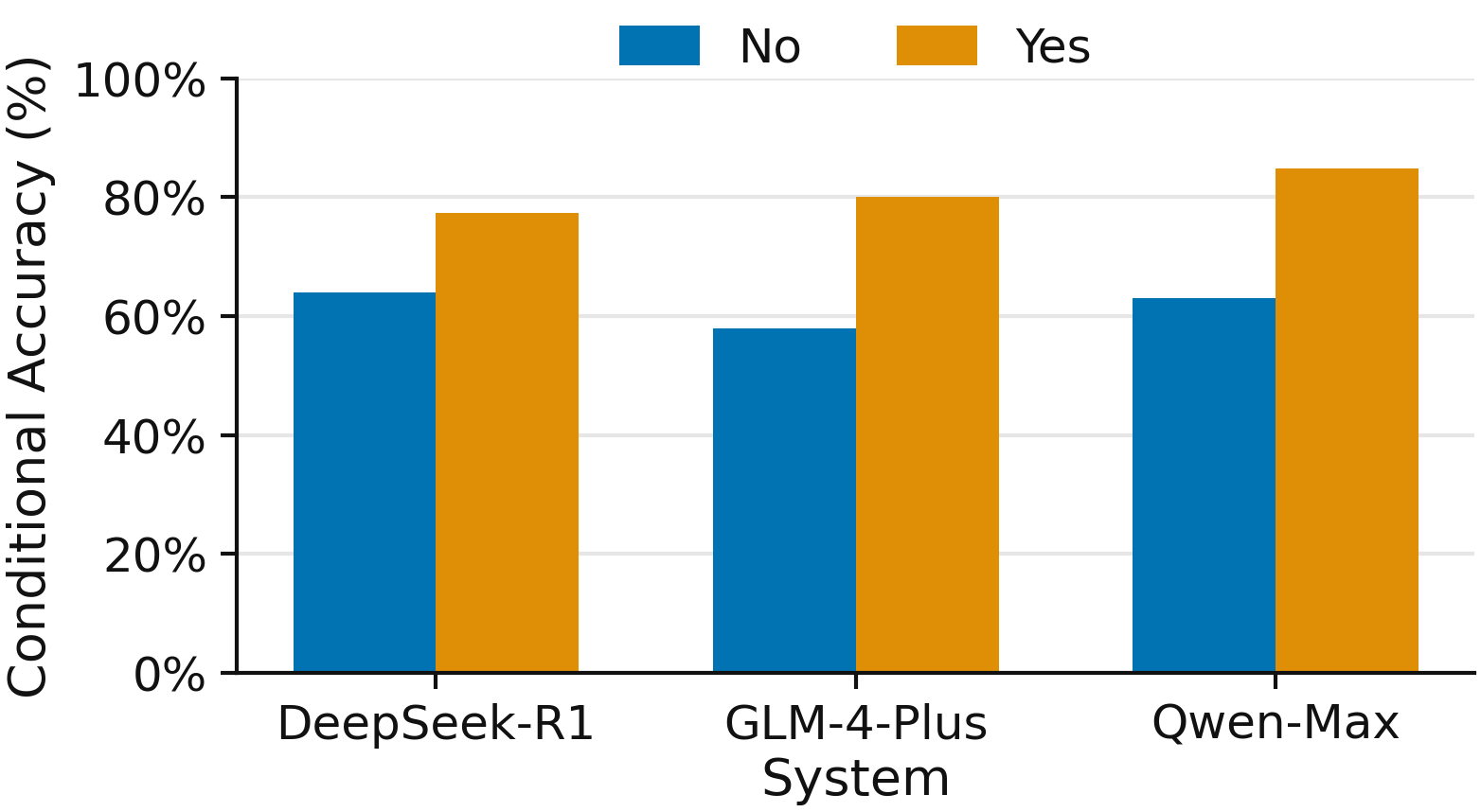}
    \caption{LLM Yes/No accuracy (excluding uncertain cases). Accuracy is computed separately for queries with Yes vs.\ No ground truth.}
    \label{fig:llm-yes-no-accuracy}
\end{figure}

Figure~\ref{fig:llm-yes-no-accuracy} shows the same polarity asymmetry for LLMs. We find that accuracy on ``Yes'' queries is consistently above 77\%, whereas accuracy on ``No'' queries falls below 65\% for all three models. 
This gap is consistent with prior work on sycophancy and affirmative bias in LLMs~\cite{hashavit2024impact}. 
Topic-level variation is visible but does not alter this pattern (see Appendix Figure~\ref{fig:appendix-llm-topic-accuracy}).

Qwen-Max and DeepSeek-R1 illustrate a tension between correctness and decision rate. Qwen-Max achieves the highest conditional accuracy (84.8\% on ``Yes'' queries), but also the highest uncertainty rate (57.3\%), meaning that it provides a definitive output for fewer than half of all queries. DeepSeek-R1 is less accurate on definitive outputs, yet provides definitive outputs for roughly three-quarters of queries (75.6\%). These findings indicate that LLM rankings depend on whether reliability is measured as correctness among definitive outputs or as the probability that a user receives a correct decision at all.

\subsection{Accuracy of AI Overviews}

\begin{figure}
    \centering
    \includegraphics[width=1\linewidth]{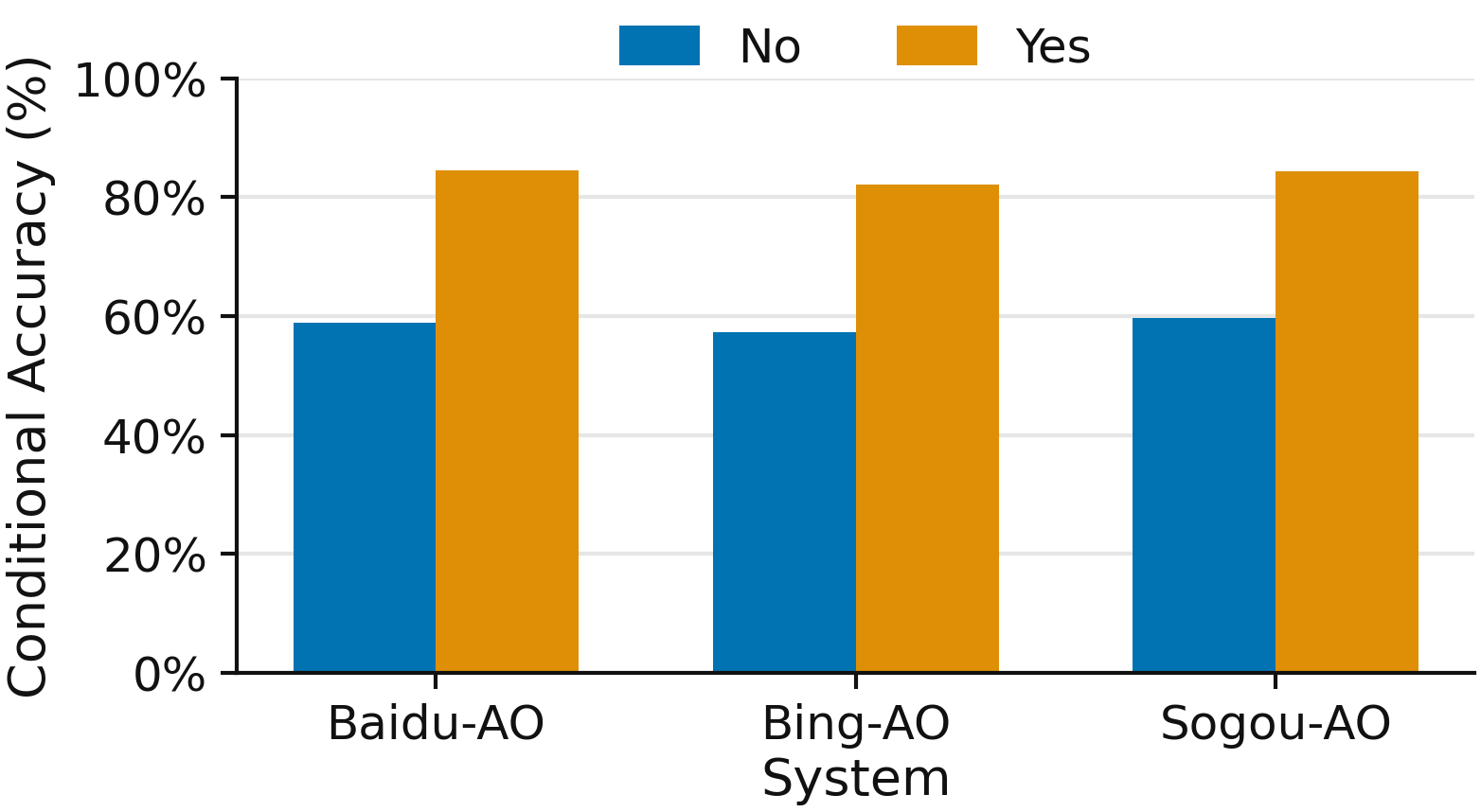}
    \caption{AI Overview Yes/No accuracy (excluding uncertain cases). Accuracy is computed separately for queries with Yes vs.\ No ground truth.}
    \label{fig:ao-yes-no-accuracy}
\end{figure}

As shown in Figure~\ref{fig:ao-yes-no-accuracy}, AI Overview systems also perform better on ``Yes'' queries than on ``No'' queries. We notice that their accuracy on ``Yes'' queries exceeds 82\%, close to the search-engine range, whereas their accuracy on ``No'' queries remains below 60\%, closer to the LLM range. Notably, AI Overviews produce uncertain outputs less often than LLMs --- roughly one in five usable outputs, compared with up to three in five for Qwen-Max --- but this higher decision rate does not yield clearly higher conditional accuracy than search engines. Topic-level differences are modest overall (see Appendix Figure~\ref{fig:appendix-ao-topic-accuracy}).

\subsection{Topic-Level Variation}

Using conditional accuracy, which excludes uncertain or no-decision outputs from the denominator, we observe topic-level variation across systems, but the largest categories do not simply determine the aggregate patterns. Health is the dominant topic in the dataset (6,610 queries), yet under the lazy search-engine model its accuracy remains in a middle band (73.0\%--74.6\%) rather than driving the aggregate as an extreme case. The clearest recurring weakness is Adult in non-LLM systems: it is the lowest topic for all three search engines (64.9\%--70.1\%) and all three AI Overviews (66.7\%--70.5\%). Since this category contains only 103 queries, these differences should be read descriptively rather than as strong topic rankings, and we avoid interpreting Adult as an intrinsically more difficult domain. Full topic-level results for each system category are reported in Appendix Figures~\ref{fig:appendix-se-topic-accuracy-lazy}, \ref{fig:appendix-se-topic-accuracy-diligent}, \ref{fig:appendix-llm-topic-accuracy}, and~\ref{fig:appendix-ao-topic-accuracy}.

\subsection{Population-Normalized Health Search Attention across Chinese Provinces}

To contextualize the potential exposure risks associated with unreliable health information, we examine where health-related search attention is more concentrated across Chinese provinces. We use Baidu Index\footnote{\url{https://index.baidu.com/v2/index.html}} as an aggregate proxy for province-level attention to health-related queries. Baidu Index provides search-volume indicators by region and time period; we collect province-level indices for health-related keywords from 2020 to 2026 and standardize province names across data sources.

Because raw search volume is strongly shaped by population size, we normalize each province's average health-related Baidu Index by its resident population from the 2020 Chinese national census. The resulting measure reflects health-related search attention per million residents. It should be interpreted as a contextual proxy for where users may be more exposed to health-related misinformation, not as a measure of system accuracy or actual individual exposure.

\begin{figure}[t]
    \centering
    \includegraphics[width=\linewidth]{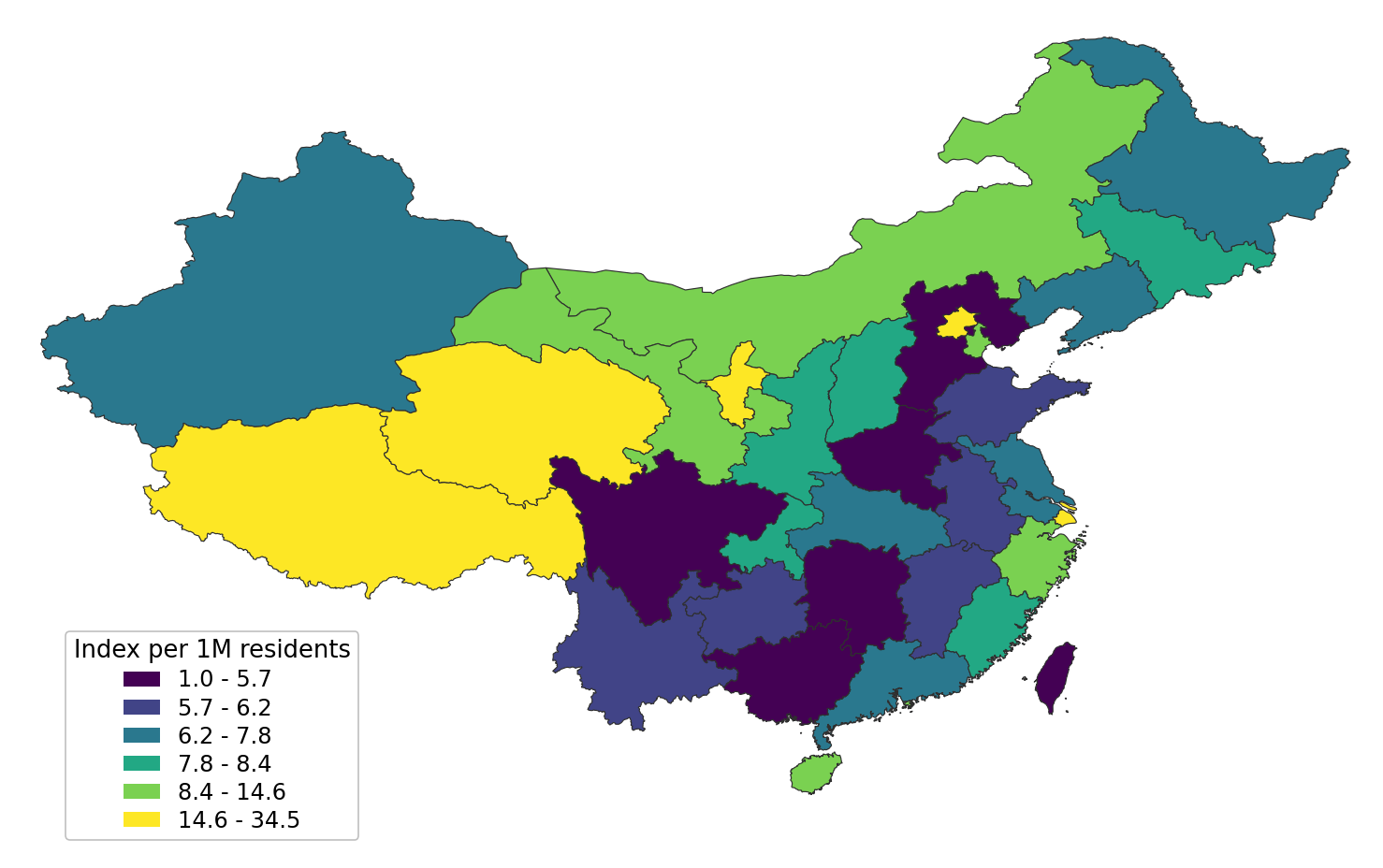}
    \caption{Population-normalized health search attention across Chinese provinces. Values reflect average health-related Baidu Index per million residents. Yellow indicates higher normalized search attention and purple indicates lower attention. The map does not encode system accuracy or measured misinformation exposure.}
    \label{fig:health-baidu-index-population-normalized}
\end{figure}

Figure~\ref{fig:health-baidu-index-population-normalized} shows that population-normalized health search attention varies substantially across provinces. Several smaller regions, including Macau, Tibet, and Qinghai, show relatively high values, while more populous provinces such as Guangdong, Jiangsu, and Shandong appear more moderate after normalization. This suggests that potential exposure to unreliable health information is not necessarily concentrated only in the most populous provinces.

We use this analysis to complement the main system-level evaluation. Since Health is the largest topic in our benchmark, the Baidu Index map helps identify geographic contexts where health information seeking is especially salient and where misinformation exposure risks may therefore deserve closer attention.

\section{Discussion and Conclusion}

With recent progress in generative AI and search-integrated answer systems, information access is increasingly shifting from traditional ranked results toward direct natural-language answers. While these systems may reduce the effort required to locate factual information, they also raise reliability concerns when users rely on direct outputs without inspecting the underlying evidence. In such settings, incorrect, uncertain, or no-decision outputs may still shape users' beliefs and decisions.

To advance our understanding of these new information-access practices, we evaluated nine systems in the Chinese web ecosystem: traditional search engines, LLMs, and search-integrated AI Overviews. Using factual Yes/No queries derived from real search logs, we compare system outputs against evidence-derived ground truth and examine correctness, decision rate, and differences between yes-labeled and no-labeled queries.

We find that conditional accuracy is broadly similar across systems once uncertain or no-decision outputs are excluded. However, systems differ substantially in decision rate: search engines provide definitive outputs most often, AI Overviews occupy an intermediate position, and LLMs produce uncertain outputs much more often. Consequently, system rankings change when uncertain or no-decision outputs are counted as errors, indicating that conditional accuracy alone provides an incomplete view of user-facing reliability.

A second finding concerns answer polarity: across search engines, LLMs, and AI Overviews, systems perform better on queries whose evidence-derived answer is \textit{Yes} than on those whose answer is \textit{No}. This gap may arise because web pages and retrieval systems may provide direct support more readily than explicit contradiction, while no-labeled queries often require evidence of absence or conflict. For LLMs, the same pattern is consistent with sycophancy or affirmative bias, where models may be more likely to accept user-framed premises than to reject them. Finally, the Baidu Index analysis shows where health-related information seeking is more concentrated after accounting for population size, helping identify contexts where exposure to unreliable health information may be more consequential.

Together, these findings show that direct-answer systems cannot be evaluated only by correctness among definitive outputs. Evaluations should also account for decision rate, performance on no-labeled queries, and demand contexts; future work should test whether retrieval-side confirmation effects and LLM sycophancy contribute to the Yes/No gap under controlled query rewrites and user-study settings.

\clearpage

\section*{Limitations}
Our study has several limitations. First, the benchmark is derived from T2Ranking search logs and therefore may not represent all Chinese information-seeking behavior across platforms, time periods, regions, or user groups. Our focus on factual Yes/No queries also excludes open-ended, explanatory, and advice-seeking information needs, although this restriction enables controlled comparison across heterogeneous systems.

Second, benchmark construction and output annotation rely partly on LLM-assisted labeling. We validate samples from each stage with human annotators, but errors may remain for ambiguous, time-sensitive, or domain-specific queries. In addition, mapping outputs to \textit{Yes}, \textit{No}, and \textit{Uncertain} simplifies richer answer behavior and does not explain why errors occur, such as whether they arise from retrieval failure, outdated evidence, hallucination, or misinterpretation. Residual model-family bias is also possible because some models used for annotation are related to models evaluated in the study.

Third, our system categories should be interpreted as user-facing access modes rather than fully separate technical architectures. Search engines expose retrieved web evidence directly, whereas AI Overviews may combine retrieval, generation, and model-internal knowledge through proprietary pipelines. Since these systems are evaluated as black boxes, our results characterize the reliability of outputs shown to users, not the mechanisms that produced them.

Fourth, our measurements capture a specific collection period. Search engines, LLMs, and AI Overview modules may change over time due to model updates, ranking changes, index updates, interface redesigns, or changes in web content. The results should therefore be interpreted as a snapshot rather than a permanent characterization of these systems.

In addition, LLMs were evaluated as black-box API systems under their default provider-side inference configurations. Because these configurations may differ across providers and are not fully controllable by users, observed differences among LLMs may partly reflect provider-specific inference settings in addition to model-level factual reliability.

Finally, the Baidu Index analysis provides population-normalized contextualization of health-related search attention, not an exposure or impact estimate. Baidu Index is a relative search-volume indicator rather than a raw query count, and population normalization does not control for internet penetration, Baidu market share, demographic composition, or healthcare access across provinces.

\section*{Ethical Considerations}

This study evaluates factual reliability using queries derived from an existing Chinese search-log benchmark and outputs collected from search engines, LLMs, and AI Overview modules. We do not collect personal information, attempt to identify users, or conduct user experiments. All results are reported in aggregate.

Because search-log queries, especially health-related queries, may reflect sensitive information needs, we treat them as aggregate information-seeking signals rather than as evidence about individual users. The Baidu Index analysis is likewise interpreted as contextual evidence of population-normalized health-related search attention across provinces; it does not estimate individual exposure, user harm, or province-specific system accuracy. The goal of this study is to support more transparent evaluation of factual reliability. It should not be interpreted as recommending any system for high-stakes factual or medical decision-making.

\noindent\textbf{Anonymous Review Artifact.}
We provide an anonymized review artifact as supplementary material, including
aggregate tables, rendered figures, and verification scripts. The artifact is
available here\footnote{\url{https://anonymous.4open.science/r/reliability-asymmetries-anonymous-artifact-859E/}}.

\bibliography{ref}

\clearpage
\appendix

\section{Additional Results}
\label{sec:appendix-additional-results}

This appendix reports additional results that support the main-text analyses. These include the full numeric summary of overall system performance, topic-level breakdowns under the conditional metric, and additional breakdowns under the metric that counts uncertain or no-decision outputs as errors among usable outputs.

\begin{table}[H]
\centering
\scriptsize
\resizebox{\columnwidth}{!}{%
\begin{tabular}{llccc}
\toprule
System & Type & Conditional Acc. & End-to-end Acc. & Uncertainty Rate \\
\midrule
Baidu Search & Search engine & 75.2 & 71.8 & 4.5 \\
Bing China & Search engine & 74.5 & 68.7 & 7.8 \\
Sogou & Search engine & 74.1 & 61.6 & 16.8 \\
\midrule
Qwen-Max & LLM & 78.9 & 33.7 & 57.3 \\
GLM-4-Plus & LLM & 73.6 & 47.9 & 34.9 \\
DeepSeek-R1 & LLM & 73.2 & 55.3 & 24.4 \\
\midrule
Baidu AI Overview & AI Overview & 73.8 & 58.4 & 20.9 \\
Bing AI Overview & AI Overview & 74.3 & 61.3 & 17.5 \\
Sogou AI Overview & AI Overview & 74.2 & 56.3 & 24.1 \\
\bottomrule
\end{tabular}
}
\caption{Full numeric summary of overall system performance. Metrics are computed over each system's usable-output set. Search-engine metrics use the lazy user model; diligent-user results are reported separately. Conditional accuracy excludes uncertain or no-decision outputs, while end-to-end accuracy counts them as errors. All values are percentages.}
\label{tab:overall-performance}
\end{table}

\begin{figure}[h]
    \centering
    \includegraphics[width=\linewidth]{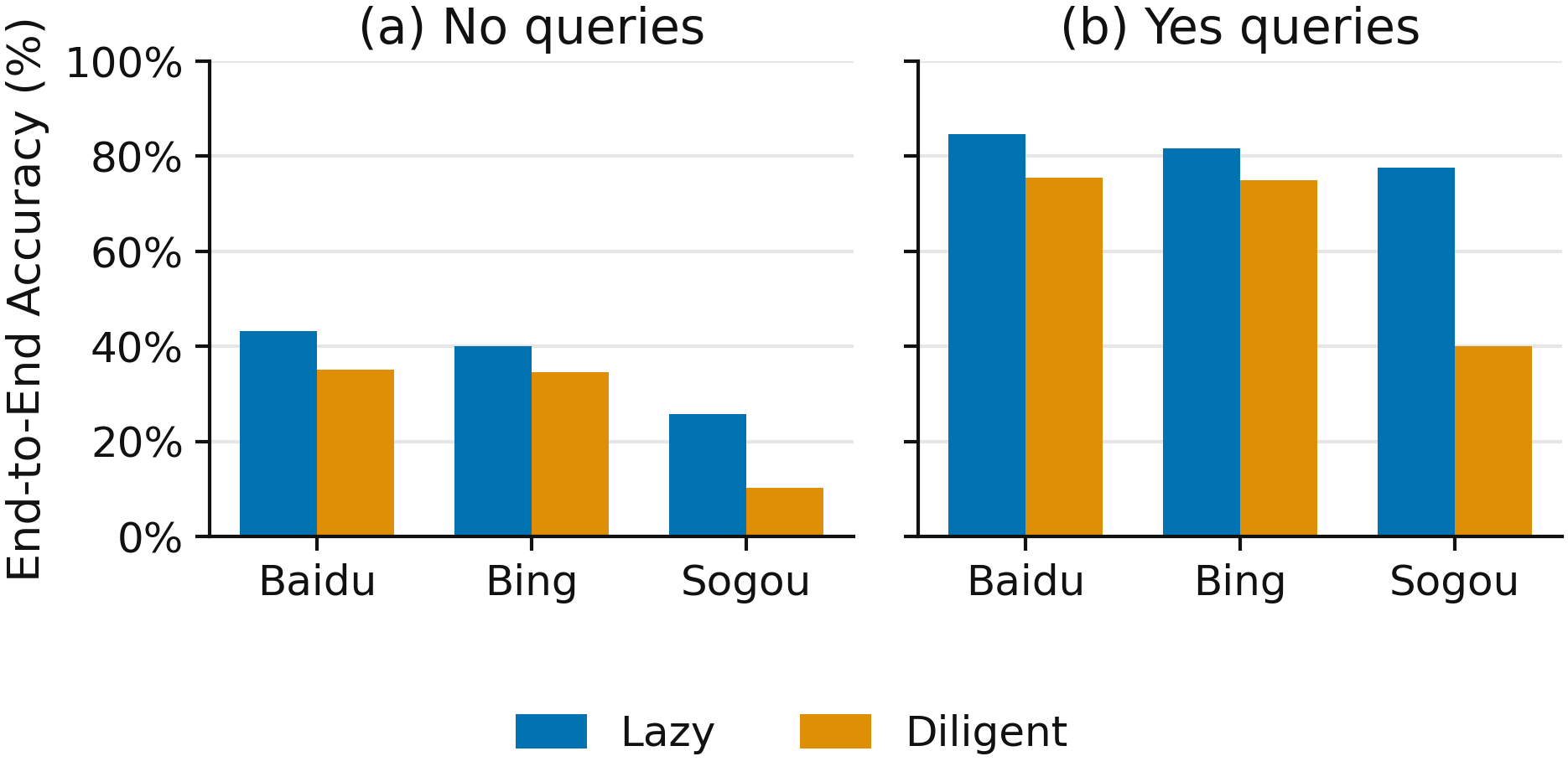}
    \caption{Search-engine accuracy by answer polarity when uncertain/no-decision cases are counted as errors. Panels show (a) No and (b) Yes ground-truth queries; bar colors compare lazy and diligent user models.}
    \label{fig:appendix-se-yes-no-accuracy-including-uncertain}
\end{figure}

\begin{figure}[H]
    \centering
    \includegraphics[width=\linewidth]{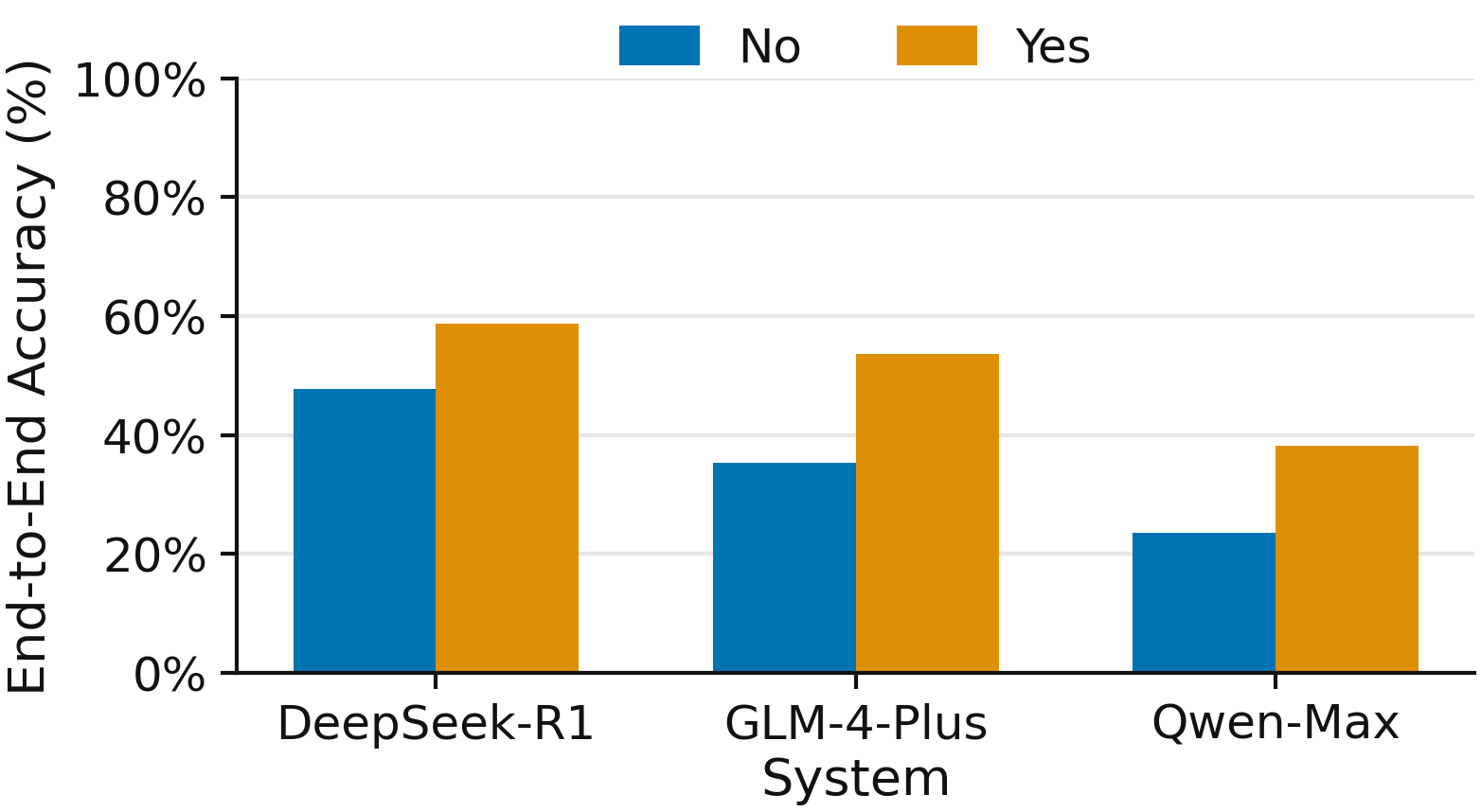}
    \caption{LLM accuracy by answer polarity when uncertain outputs are counted as errors.}
    \label{fig:appendix-llm-yes-no-accuracy-including-uncertain}
\end{figure}

\begin{figure}[H]
    \centering
    \includegraphics[width=\linewidth]{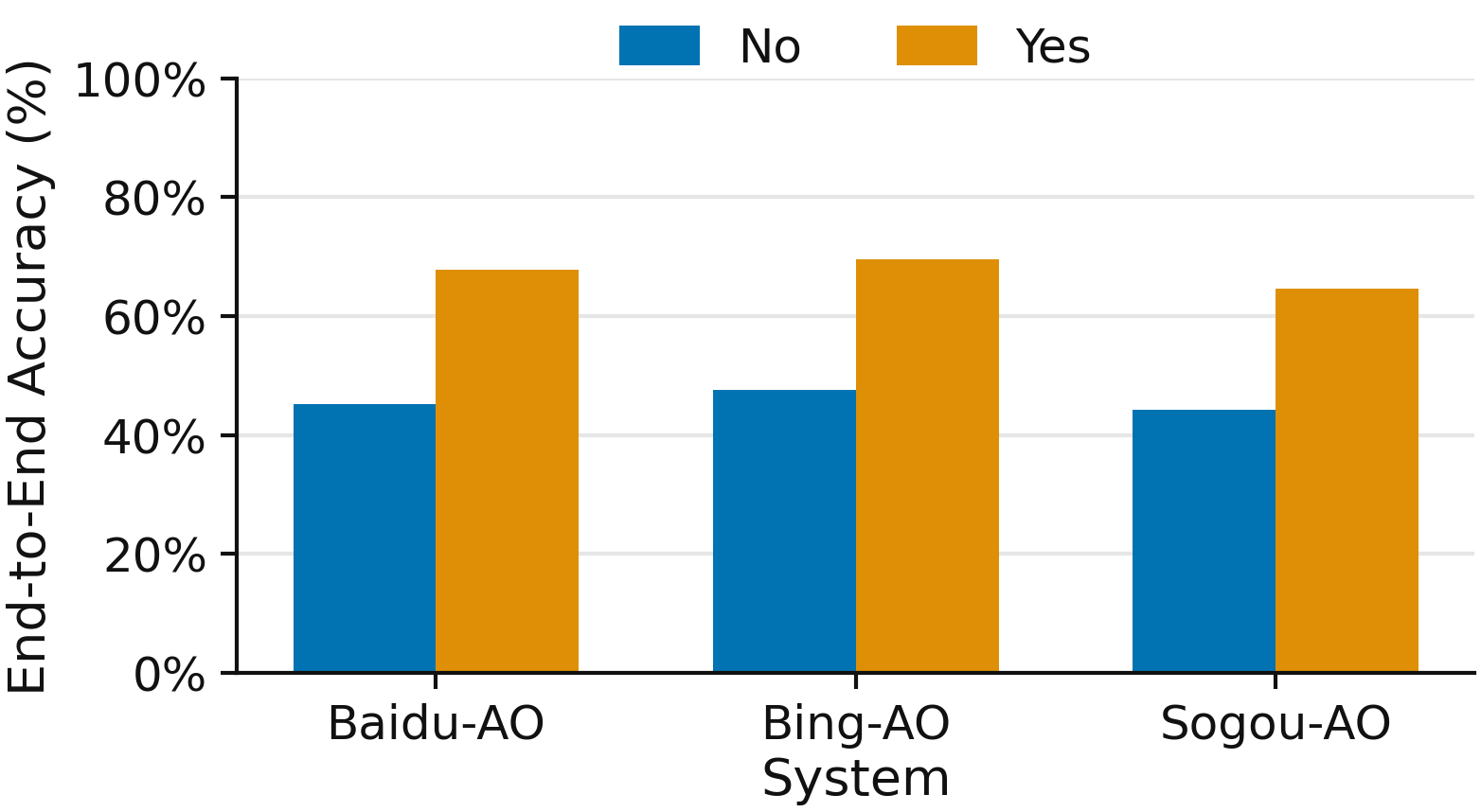}
    \caption{AI Overview accuracy by answer polarity when uncertain outputs are counted as errors.}
    \label{fig:appendix-ao-yes-no-accuracy-including-uncertain}
\end{figure}

\begin{figure*}[t]
    \centering
    \includegraphics[width=\textwidth]{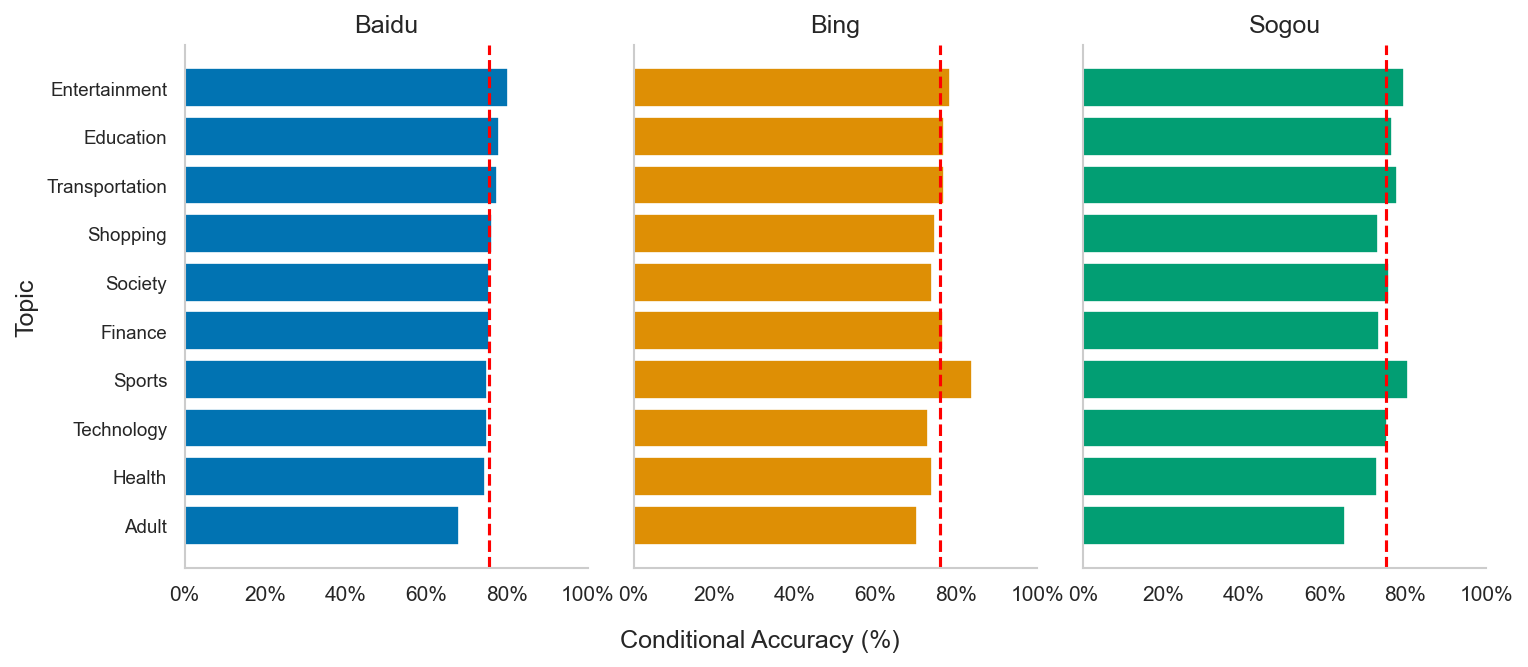}
    \caption{Search engine topic-level accuracy (lazy user model, excluding no-decision cases). Each panel shows accuracy by topic for one search engine, sorted by accuracy. The dashed vertical line indicates the mean accuracy across topics.}
    \label{fig:appendix-se-topic-accuracy-lazy}
\end{figure*}

\begin{figure*}[t]
    \centering
    \includegraphics[width=\textwidth]{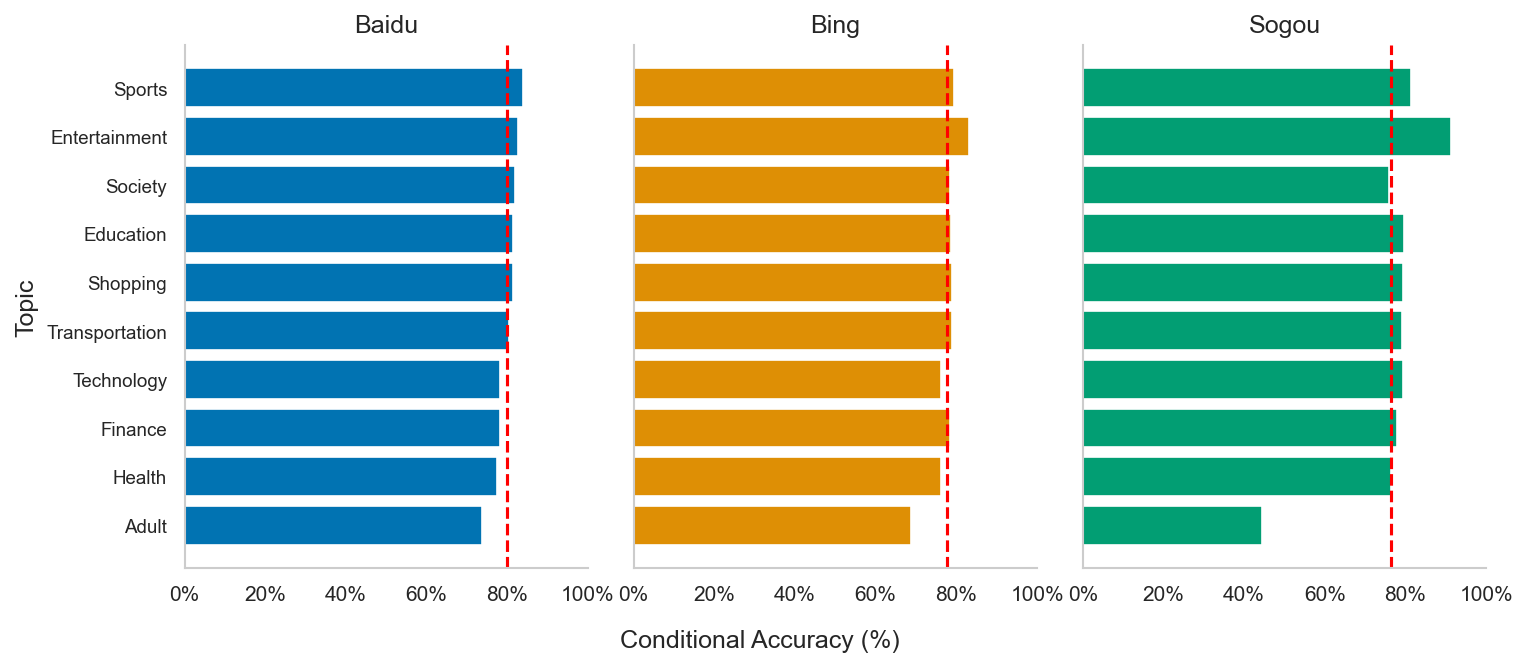}
    \caption{Search engine topic-level accuracy (diligent user model, excluding no-decision cases). Each panel shows accuracy by topic for one search engine, sorted by accuracy. The dashed vertical line indicates the mean accuracy across topics.}
    \label{fig:appendix-se-topic-accuracy-diligent}
\end{figure*}

\begin{figure*}[t]
    \centering
    \includegraphics[width=\textwidth]{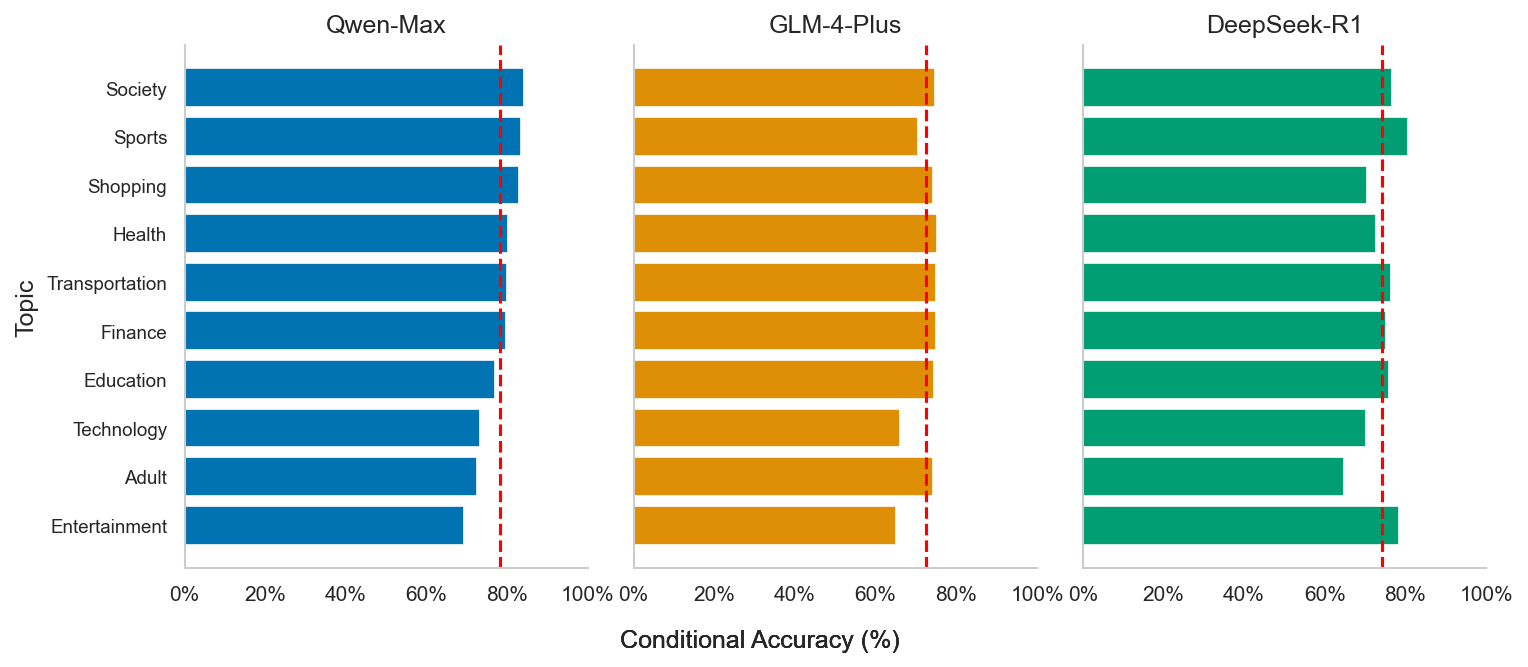}
    \caption{LLM topic-level accuracy (excluding uncertain cases). Each panel shows accuracy by topic for one LLM, sorted by accuracy. The dashed vertical line indicates the mean accuracy across topics.}
    \label{fig:appendix-llm-topic-accuracy}
\end{figure*}

\begin{figure*}[t]
    \centering
    \includegraphics[width=\textwidth]{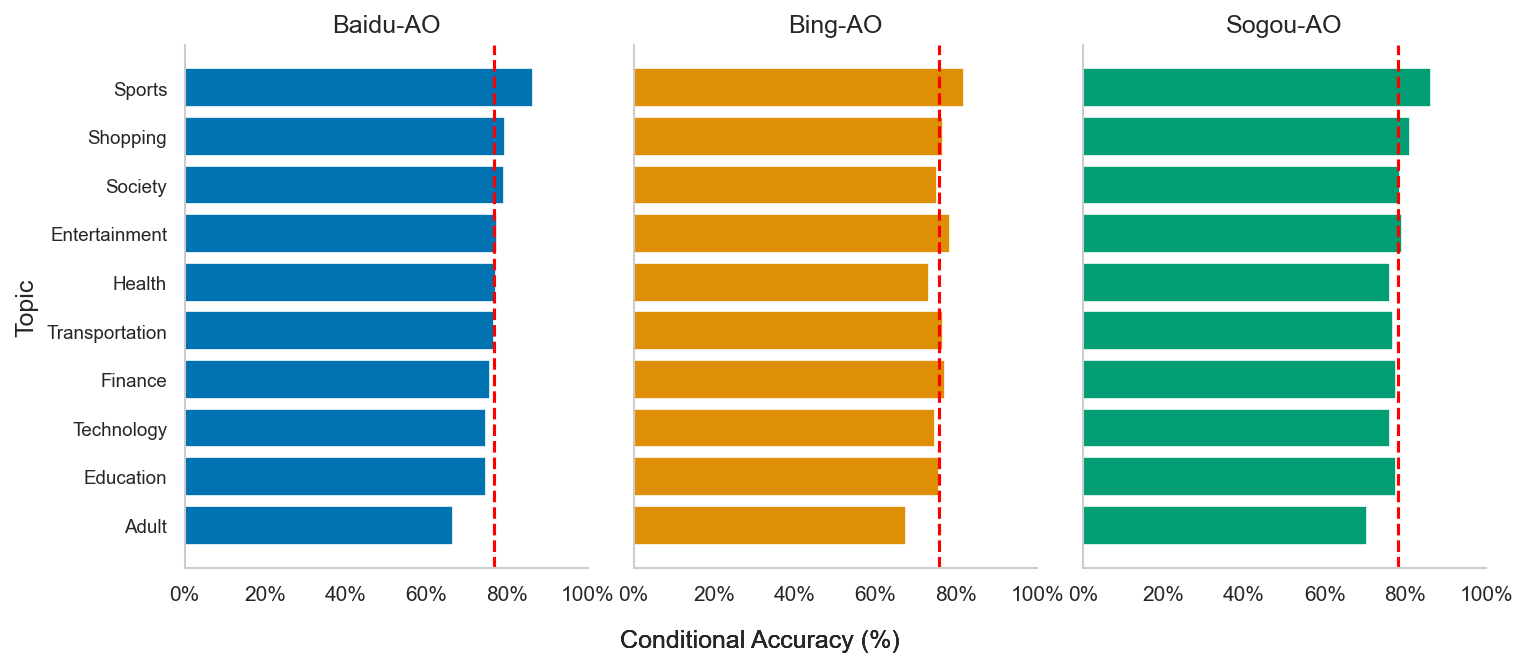}
    \caption{AI Overview topic-level accuracy (excluding uncertain cases). Each panel shows accuracy by topic for one AI Overview system, sorted by accuracy. The dashed vertical line indicates the mean accuracy across topics.}
    \label{fig:appendix-ao-topic-accuracy}
\end{figure*}

\begin{figure*}[t]
    \centering
    \includegraphics[width=\textwidth]{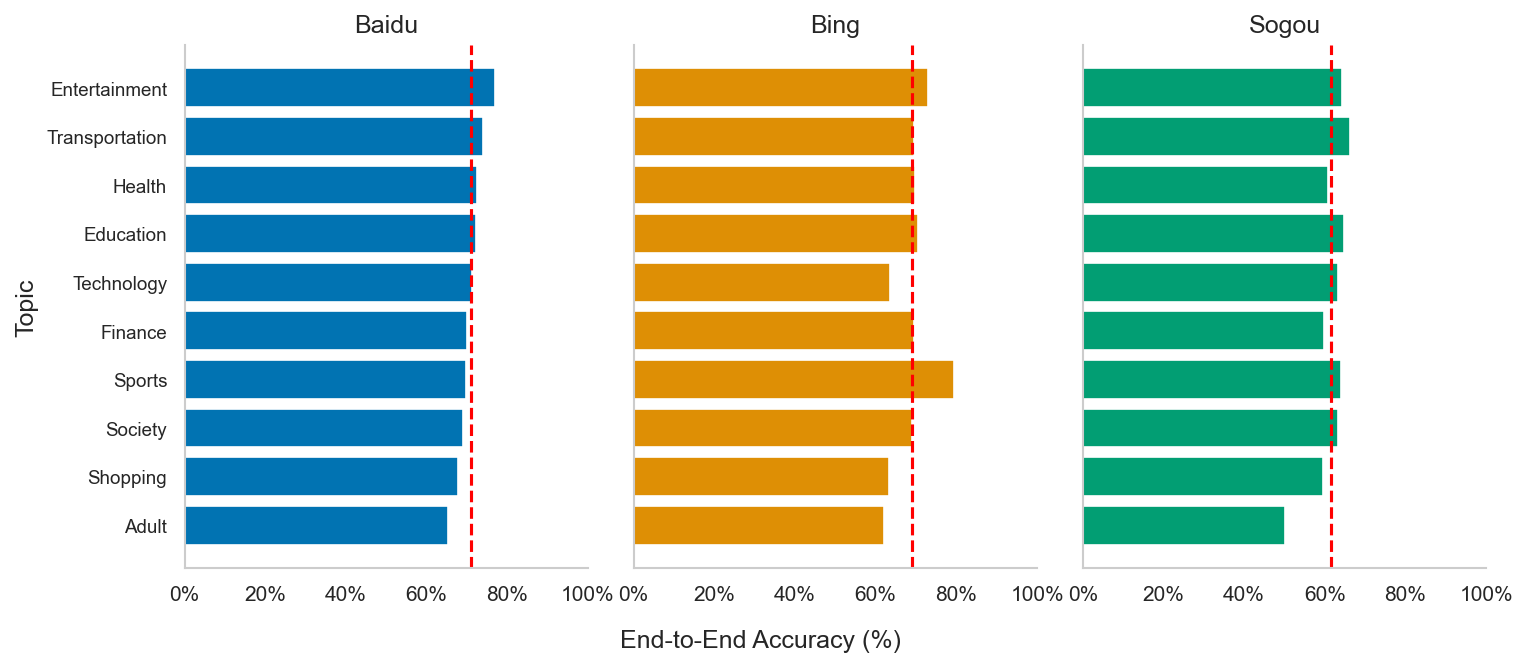}
    \caption{Search-engine topic-level accuracy including uncertain/no-decision as error under the lazy user model. Each panel corresponds to one search engine, with topics sorted by accuracy. The dashed vertical line indicates panel-wise mean accuracy.}
    \label{fig:appendix-se-topic-accuracy-lazy-including-uncertain}
\end{figure*}

\begin{figure*}[t]
    \centering
    \includegraphics[width=\textwidth]{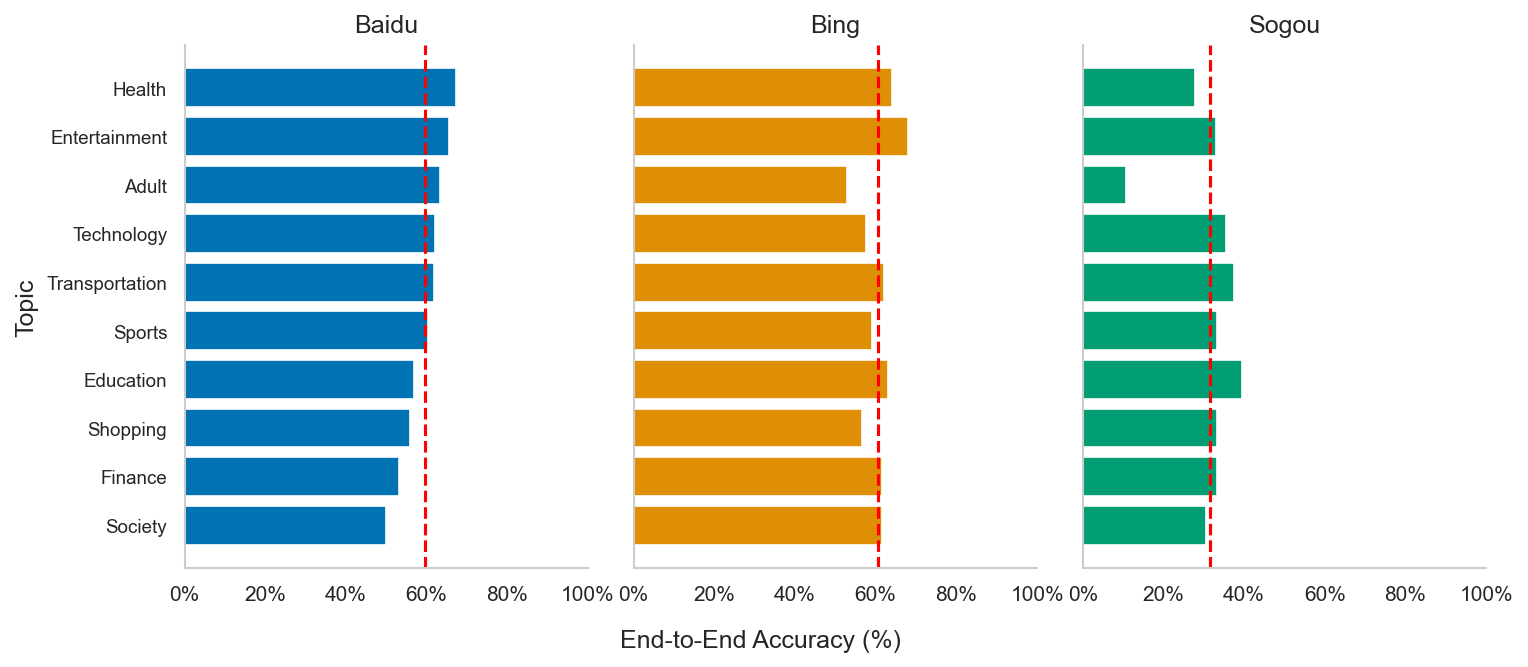}
    \caption{Search-engine topic-level accuracy including uncertain/no-decision as error under the diligent user model. Each panel corresponds to one search engine, with topics sorted by accuracy. The dashed vertical line indicates panel-wise mean accuracy.}
    \label{fig:appendix-se-topic-accuracy-diligent-including-uncertain}
\end{figure*}

\begin{figure*}[t]
    \centering
    \includegraphics[width=\textwidth]{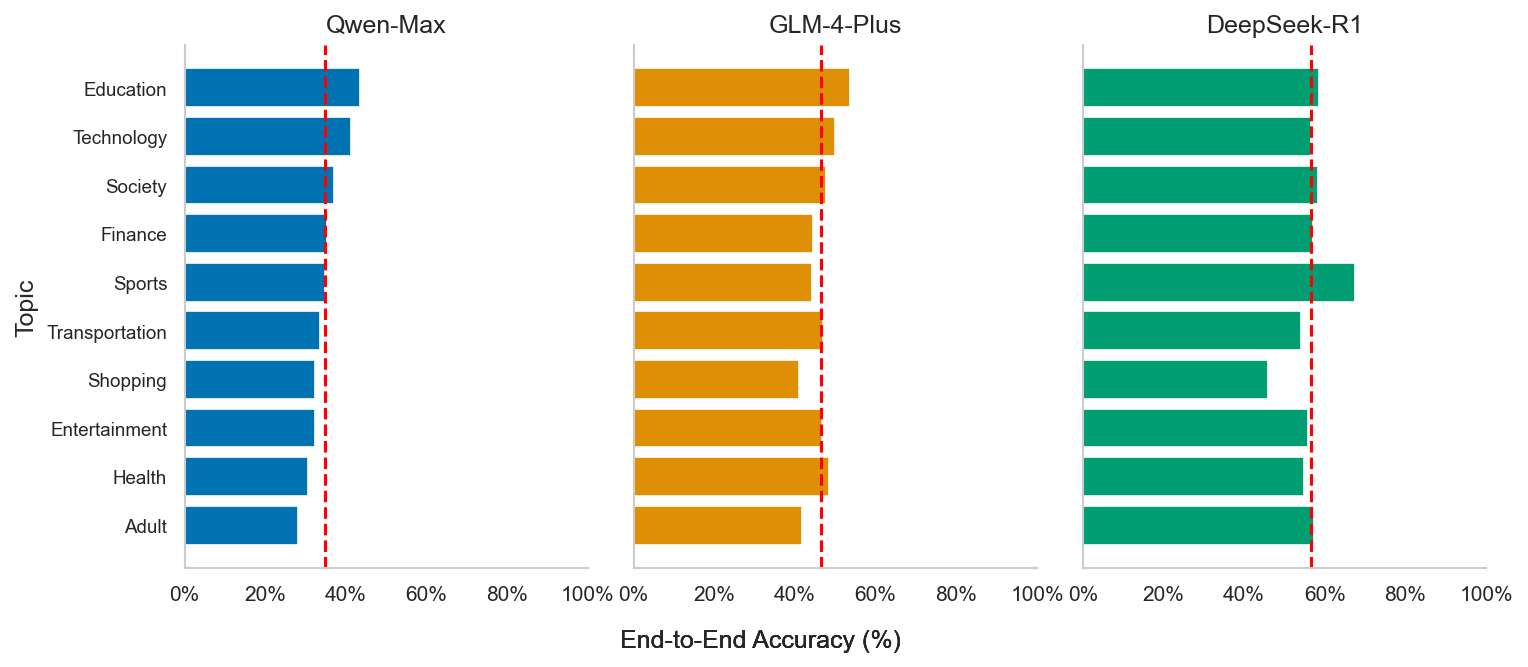}
    \caption{LLM topic-level accuracy including uncertain as error. Each panel corresponds to one model, with topics sorted by accuracy. The dashed vertical line indicates panel-wise mean accuracy.}
    \label{fig:appendix-llm-topic-accuracy-including-uncertain}
\end{figure*}

\begin{figure*}[t]
    \centering
    \includegraphics[width=\textwidth]{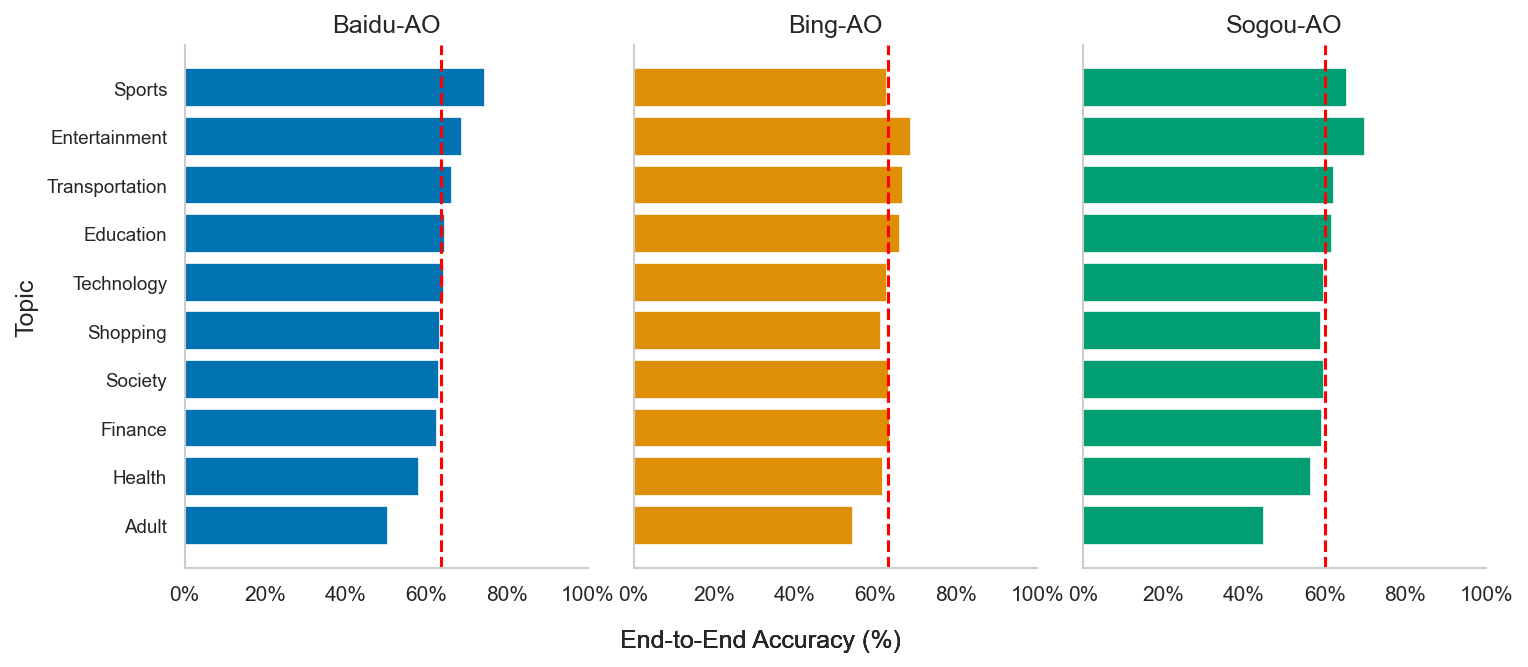}
    \caption{AI Overview topic-level accuracy including uncertain as error. Each panel corresponds to one AI Overview system, with topics sorted by accuracy. The dashed vertical line indicates panel-wise mean accuracy.}
    \label{fig:appendix-ao-topic-accuracy-including-uncertain}
\end{figure*}

\clearpage

\section{Prompt Templates}
\label{app:prompts-bilingual}

\subsection{System-Output Evaluation Prompts}

\paragraph{A. Search Engine Passage Evaluation Prompt (EN)}
\begin{quote}\small
You are a fact-checking assistant.

You will receive:
(1) a user query, and
(2) a web passage retrieved by a search engine.

Your task is to determine whether the passage provides a factual answer to the query.

Decision rules:
- Output \textbf{Yes} if the passage clearly supports the proposition in the query.
- Output \textbf{No} if the passage clearly contradicts the proposition.
- Output \textbf{Uncertain} if the passage does not provide enough information, is irrelevant, or does not directly answer the query.

Important constraints:
- Use only the provided passage.
- Do not use background knowledge.
- Do not infer missing information.
- Return exactly one token: \textbf{Yes / No / Uncertain}.

Query: \{query\}

Web passage: \{web\_content\}
\end{quote}

\begin{CJK*}{UTF8}{gbsn}
\paragraph{A. 搜索引擎网页片段判定提示词（中文）}
\begin{quote}\small
你是一名事实核查助手。

你将收到：
（1）一个用户问题；
（2）一段由搜索引擎检索得到的网页内容。

你的任务是判断该网页内容是否对该问题提供了事实性回答。

判定规则：
- 若内容明确支持问题中的命题，输出\textbf{Yes}；
- 若内容明确反驳问题中的命题，输出\textbf{No}；
- 若内容信息不足、与问题无关、或未直接回答问题，输出\textbf{Uncertain}。

重要约束：
- 只能依据提供的网页内容判断；
- 不得使用背景知识；
- 不得推断缺失信息；
- 只输出一个词：\textbf{Yes / No / Uncertain}。

问题：\{query\}

网页内容：\{web\_content\}
\end{quote}
\end{CJK*}

\paragraph{B. AI Overview Evaluation Prompt (EN)}
\begin{quote}\small
You are a fact-checking assistant.

You will receive a user query and an AI overview generated by a search engine.
Your task is to determine whether the overview provides a factual answer.

Decision rules:
- Output \textbf{Yes} if the overview clearly supports the proposition.
- Output \textbf{No} if the overview clearly contradicts the proposition.
- Output \textbf{Uncertain} if the overview is insufficient, irrelevant, or does not directly answer the query.

Important constraints:
- Use only the provided overview text.
- Do not use background knowledge.
- Do not infer missing information.
- Return exactly one token: \textbf{Yes / No / Uncertain}.

Query: \{query\}

Overview: \{overview\_text\}
\end{quote}

\begin{CJK*}{UTF8}{gbsn}
\paragraph{B. AI Overview 判定提示词（中文）}
\begin{quote}\small
你是一名事实核查助手。

你将收到一个用户问题，以及一个由搜索引擎生成的 AI Overview 回答。
你的任务是判断该 Overview 是否对问题提供了事实性回答。

判定规则：
- 若 Overview 明确支持或肯定问题中的命题，输出\textbf{Yes}；
- 若 Overview 明确反驳或否定问题中的命题，输出\textbf{No}；
- 若 Overview 信息不足、与问题无关、或未直接回答问题，输出\textbf{Uncertain}。

重要约束：
- 只能依据提供的 Overview 文本判断；
- 不得使用背景知识；
- 不得推断缺失信息；
- 只输出一个词：\textbf{Yes / No / Uncertain}。

问题：\{query\}

Overview 回答：\{overview\_text\}
\end{quote}
\end{CJK*}

\paragraph{C. LLM Native Zero-Context Prompt (EN)}
\begin{quote}\small
You are a fact-checking assistant.

Answer the following factual Yes/No query based only on your internal knowledge.
Do not search the web. Do not provide explanations.

Decision rules:
- \textbf{Yes}: the proposition is factually supported.
- \textbf{No}: the proposition is factually contradicted.
- \textbf{Uncertain}: insufficient confidence, ambiguity, or lack of reliable information.

Return exactly one token: \textbf{Yes / No / Uncertain}.

Query: \{query\}
\end{quote}

\begin{CJK*}{UTF8}{gbsn}
\paragraph{C. LLM 原生零上下文判定提示词（中文）}
\begin{quote}\small
你是一名事实核查助手。

请仅基于你自身已有知识回答下面的事实性是/否问题。
不要搜索网页，不要输出解释。

判定规则：
- \textbf{Yes}：命题在事实上成立或被支持；
- \textbf{No}：命题在事实上不成立或被否定；
- \textbf{Uncertain}：把握不足、问题存在歧义，或缺乏可靠信息。

只输出一个词：\textbf{Yes / No / Uncertain}。

问题：\{query\}
\end{quote}
\end{CJK*}

\subsection{Dataset Construction Prompts}

\subsubsection{Yes-No Question Identification Prompt}
\label{appendix:yes_no_question} 

\begin{CJK*}{UTF8}{gbsn} 
请判断以下输入是否为``是/否问题”。  

\textbf{定义：}  
是/否问题是一类可以用简单的``是”或``否”作答的问题。  

\textbf{任务：}  
给定一个输入，判断其是否为是/否问题。  

\textbf{输入示例与输出：}  
- 输入：39.5是发烧吗？  
  输出：yes  

- 输入：枯藤是生物吗  
  输出：yes  

- 输入：bb霜是粉底液吗  
  输出：yes  

- 输入：久咳嗽不好是什么原因  
  输出：No  

- 输入：tfboys为什么现在不能提及对方  
  输出：No  

- 输入：咳嗽怎么睡觉不咳嗽  
  输出：No  

\textbf{输出要求：}  
仅返回``yes”或``No”，不作其他解释。  
\end{CJK*} 

\vspace{0.5em} 
\noindent\textbf{English Version:} \\  
This prompt asks the model to identify whether a given query is a yes-no question.  

\textbf{Definition:}  
A yes-no question is one that can be answered with a simple \textit{``Yes''} or \textit{``No''}.  

\textbf{Task:}  
Given a query, determine whether it qualifies as a yes-no question.  

\textbf{Examples:}  
\begin{CJK*}{UTF8}{gbsn} 
\begin{itemize} 
    \item Input: ``39.5是发烧吗？'' → Output: \textit{yes}  
    \item Input: ``枯藤是生物吗'' → Output: \textit{yes}  
    \item Input: ``bb霜是粉底液吗'' → Output: \textit{yes}  
    \item Input: ``久咳嗽不好是什么原因'' → Output: \textit{No}  
    \item Input: ``tfboys为什么现在不能提及对方'' → Output: \textit{No}  
    \item Input: ``咳嗽怎么睡觉不咳嗽'' → Output: \textit{No} 
\end{itemize} 
 \end{CJK*} 
\textbf{Expected Output:}  
Only return \textit{``yes''} or \textit{``No''}, without additional explanation. 

\subsubsection{Answer Labeling Prompt}
\label{appendix:answer_labeling_prompt} 
\begin{CJK*}{UTF8}{gbsn} 
你是一名事实核查助手。现在给定一个用户提出的问题，以及与该问题相关的网页内容（片段）。你的任务是根据该网页内容判断网页是否给出了明确的回答。 

\textbf{判断标准：} 

选择 ``是”：如果网页内容明确支持、证实或肯定问题中的陈述。 

选择 ``否”：如果网页内容明确反对、否定或与问题中的陈述相矛盾。 

选择 ``不确定”：如果网页内容未包含足够信息来判断真假，或内容与问题无关。 

\textbf{注意事项：} 

必须严格依据网页内容做判断，不得使用你的背景知识。 

只返回以下三个词之一： 
是 / 否 / 不确定 

问题： {query} 
网页内容： {passage} 

请根据上述标准作答。 
\end{CJK*} 
\vspace{0.5em} 
\textbf{English Version:} 

You are a fact-checking assistant. You are given a user question and a passage retrieved from the web. Your task is to determine whether the passage provides a factual answer to the question. 

Decision Criteria: 

Output ``Yes” if the passage clearly supports or affirms the proposition in the query. 

Output ``No” if the passage clearly contradicts or refutes the proposition. 

Output ``Uncertain” if the passage does not contain enough information to determine a factual answer or is unrelated to the query. 

Important: 
You must rely only on the provided passage, not on prior knowledge. 
Return only one word: Yes / No / Uncertain. 

Question: {query} 
Passage: {passage} 

\subsubsection{Fact-Checking Question Classification Prompt}
\label{appendix:fact_checking} 

\begin{CJK*}{UTF8}{gbsn} 
请判断给定的是/否问题是否属于 fact-checking 问题。  

\textbf{定义：}  
事实（fact）是关于现实世界的客观陈述，可以通过证据、观察或权威来源独立验证，独立于个人观点、感受或信仰。  
fact-checking question 是指答案指向事实的是/否问题，其内容能够被客观地验证为真或假，可借助外部证据、科学知识、历史记录或权威文件加以确认。  

\textbf{任务：}  
对于输入的是/否问题，判断其是否为 fact-checking 问题。  

\textbf{判定标准：}  
- 若问题的答案可以通过事实加以验证（客观可判定真/假），输出：``fact-checking”。  
- 若问题依赖于个人感受、主观经验、个体差异或特定情境，输出：``not fact-checking”。  

\textbf{输入示例：}  
``\{输入问题\}''  

\textbf{输出要求：}  
仅输出一个标签：``fact-checking” 或 ``not fact-checking”。 
\end{CJK*} 

\vspace{0.5em} 
\noindent\textbf{English Version:} \\  
The following prompt was used to classify Yes/No questions as fact-checking or not.  

\textbf{Definition:}  
A \textit{fact} is defined as an objective statement about the real world that can be independently verified through evidence, observation, or authoritative sources, and is independent of personal opinions, feelings, or beliefs.  
A \textit{fact-checking question} is a Yes/No question whose answer refers to factual content and can be objectively verified as true or false by means of external evidence, scientific knowledge, historical records, or authoritative documents.  

\textbf{Task:}  
Given a Yes/No question, determine whether it qualifies as a fact-checking question.  

\textbf{Decision Criteria:}  
\begin{itemize} 
    \item If the answer can be verified as true or false based on factual evidence, output: \textit{``fact-checking''}.  
    \item If the question depends on personal feelings, subjective experience, individual differences, or context-specific conditions, output: \textit{``not fact-checking''}.  
\end{itemize} 

\textbf{Expected Output:}  
A single label: \textit{``fact-checking''} or \textit{``not fact-checking''}. 

\subsubsection{Topic Recognition Prompt}
\label{appendix:topic_recognition} 

\begin{CJK*}{UTF8}{gbsn} 
请判断以下句子的主题，并从给定的标签中选择最合适的一项： \\ 

可选标签：  

\begin{itemize} 
    \item 健康 (Health)  
\item 教育 (Education)  
\item 科技 (Technology)  
\item 娱乐 (Entertainment)  
\item 体育 (Sports)  
\item 购物 (Shopping)  
\item 出行 (Travel/Transportation)  
\item 财经 (Finance/Business)  
\item 社会 (Society/Politics/Law)  
\item 成人 (Adult)  
\end{itemize} 

如果无法判断，请回答：无法确定。 \\ 

问题：``{输入问题}'' 
\end{CJK*} 

\vspace{0.5em} 
\noindent\textbf{English Description:} \\ 
This prompt asks the model to determine the topic of a given Chinese sentence. 
The model must select one label from the predefined list (included in the prompt). 
If the topic cannot be determined with confidence, the model should respond with 
\textit{``unable to determine''}.

\end{document}